\documentclass[10pt]{article}

\usepackage[margin=0.75in]{geometry}
\usepackage{amsmath,amssymb}
\usepackage{graphicx}
\usepackage{bm}
\usepackage{hyperref}
\usepackage{caption}
\usepackage{subcaption}
\usepackage{microtype}

\usepackage{bibunits}
\defaultbibliography{references}
\defaultbibliographystyle{unsrt}

\usepackage{titlesec}
\titleformat{\section}
  {\normalfont\large\bfseries}
  {}{0pt}{}
\titleformat{\subsection}
  {\normalfont\normalsize\bfseries}
  {}{0pt}{}
\titlespacing*{\section}{0pt}{2ex}{0.8ex}

\title{\textbf{Brush-mediated angular constraints reshape structure, rigidity, and percolation in colloidal depletion gels}}

\author{
Ziye Zhuang$^{1,*}$,
Robert A. Campbell$^{2,*}$,
Paniz Haghighi$^{2}$,
Safa Jamali$^{2,3,\dagger}$,
Ali Mohraz$^{1,\dagger}${\vspace{0.2cm}}\\
{\small
$^{1}$Department of Chemical and Biomolecular Engineering, University of California, Irvine, CA$,$ 92697, USA}\\
{\small $^{2}$Department of Mechanical and Industrial Engineering$,$ Northeastern University$,$ Boston$,$ MA$,$ 02115$,$ USA}\\
{\small $^{3}$Department of Chemical Engineering$,$ Northeastern University$,$ Boston$,$ MA$,$ 02115$,$ USA}{\vspace{0.2cm}}\\
{\small $^{*}$These authors contributed equally.}\\
{\small $^{\dagger}$Correspondence: s.jamali@northeastern.edu, mohraz@uci.edu}
}
\date{}

\begin{document}


\twocolumn[
\begin{@twocolumnfalse}

\maketitle

\begin{abstract}
Colloidal gels, like many other soft and disordered solids derive their mechanical properties not only from the strength of interparticle attraction, but also from the symmetry of the forces that constrain particle motion. While non-central interactions are known to profoundly alter rigidity and elasticity, they are typically introduced through particle anisotropy, surface roughness, or patchy interactions, obscuring their independent role. Here we demonstrate a minimal and geometry-preserving route to emergent non-central forces in colloidal gels by reducing the density of surface-grafted polymer brushes. At low brush density, partial brush interpenetration introduces an effective angular bending rigidity at particle contacts, despite fully isotropic particle geometry. This emergent constraint suppresses local densification, stabilizes low-coordination networks, and produces highly ramified gel structures with enhanced elasticity. Combining experiments, simulations, and mean-field theory, we show that these non-central constraints reorganize structure and mechanics across length scales, shifting gelation boundaries and increasing the elastic modulus by nearly a factor of three. Our results establish surface brush density as a generic control parameter for programming interaction symmetry in soft particulate matter, with implications for rigidity, percolation, and mechanical design in disordered systems.
\end{abstract}

\vspace{0.8cm}

\end{@twocolumnfalse}
]

\begin{bibunit}

\section{Main}

Colloidal gels represent a ubiquitous class of soft materials used in wide ranging applications such as membranes, food, personal care products, coatings, and additive manufacturing \cite{liang_stiff_2025,xiong_dynamic_2019,zhang_supramolecular_2025}. This broad utility stems in part from their rich viscoelastic properties. Namely, colloidal gels exhibit a weak elastic modulus due to a sample-spanning network of weakly attractive particles that can break down in response to applied deformation.  \cite{colombo_stress_2014,landrum_delayed_2016,winter_analysis_1986,studart_yielding_2011,muller_toughening_2023}. The nature of this viscoelastic response is strongly dependent on the underlying particle network structure and the details of interparticle interactions \cite{whitaker_colloidal_2019,nabizadeh_network_2024,mangal_predicting_2024,zaccone_elasticity_2009,bantawa_hidden_2023}. 

A versatile method for prompting colloidal aggregation and gel network formation is the addition of non-adsorbing species such as polymers, which induce attractive depletion interactions between colloids \cite{asakura_interaction_1958,vrij_polymers_1976,dibble_structure_2006}. The microstructure and rheology of the resulting suspension are particularly sensitive to the details of these depletion interactions. Previous research has demonstrated microstructures ranging from bicontinuous morphologies reminiscent of arrested spinodal decomposition to tenuous networks described by fractal geometry, each with their own rheological signatures \cite{lu_gelation_2008, van_schooneveld_structure_2009, nabizadeh_network_2024}. This microstructural diversity has been achieved by adjusting the size and concentration of the depletant species, which in turn set the range and strength of attractive interparticle forces, respectively \cite{lu_gelation_2008,lu__colloidal_2013}. 

\begin{figure*}[ht!]
\centering
\includegraphics[width=1\linewidth]{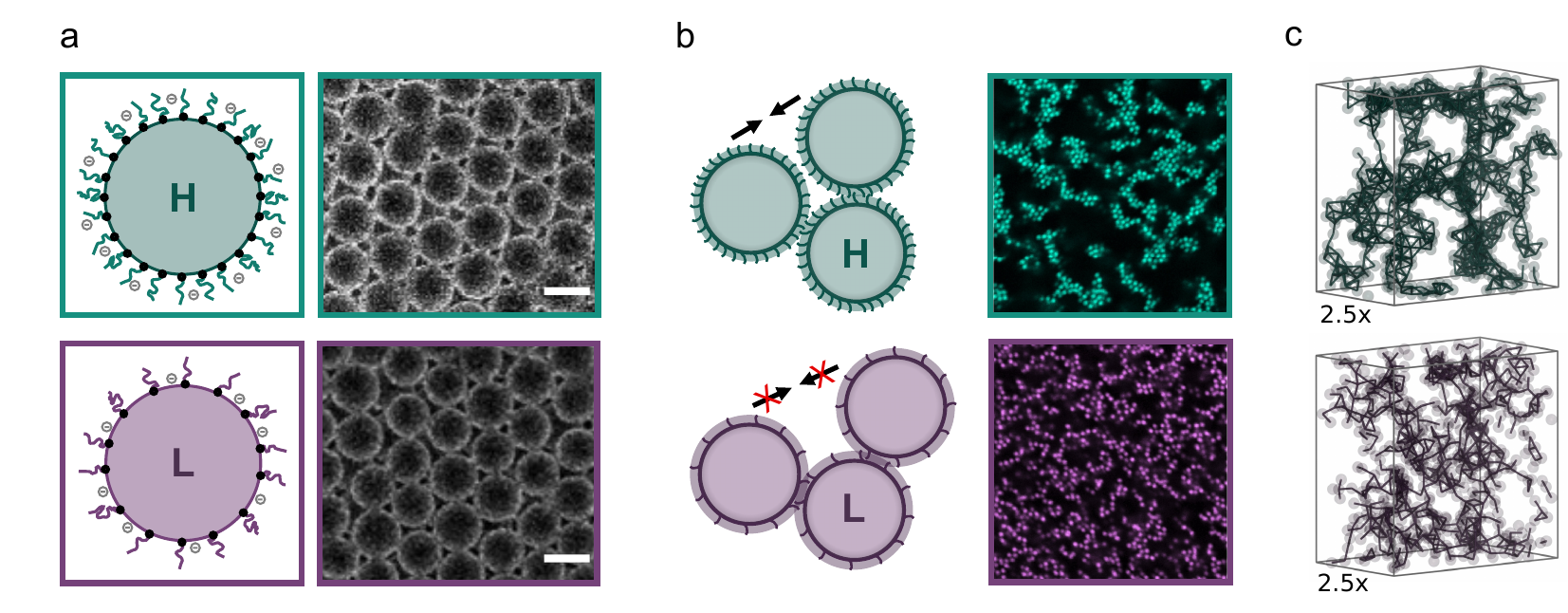}
\caption{\textbf{Reducing brush-density introduces non-central constraints without changing particle geometry.} (a) Brush density is reduced by decreasing the amount of 2-(2-bromoisobutyryloxy) ethyl acrylate inimer (black) and the total monomer feed for DMA-\textit{co}-SPAm brush growth (color), while preserving particle size and morphology (SEM; scale bar: 1$\mu m$). (b) High brush density (top) favors compact, cluster-rich assembly, whereas lower brush density (bottom) introduces brush-mediated constraints that prevent compact clustering and yield thinner, more ramified gel strands. Confocal snapshots (right) show the corresponding microstructure under $c/c^* = 0.6$ and 10 mM NaCl and at the total volume fraction of $\phi=0.1$. (c) 3D network reconstructions from experimental z-stacks ($c/c^* = 0.6$, 10 mM NaCl). Lines indicate interparticle bonds and transparent spheres show particle positions.}
\label{fig:summary}
\end{figure*}

Fundamentally, depletion interactions result from the size-exclusion of non-adsorbing species from a shell surrounding each colloidal particle \cite{asakura_interaction_1958, louis_polymer_2002}. As such, the geometry of these interactions is assumed to mimic that of the primary particles, resulting in centrosymmetric forces in suspensions of colloidal spheres \cite{yang_depletion_2006}. Recent experimental and computational studies have shown that interfacial friction, interlocking geometries, and surface topology can introduce non-central forces in a colloidal system and significantly alter this picture \cite{hsiao_rheological_2017,hsu_roughness-dependent_2018,scherrer_characterizing_2025, wang_surface_2019, nguyen_computer_2020, immink_using_2020}. For example, the introduction of measurable surface roughness and interlocking geometries can change the gelation boundary and bulk rheology of colloidal suspensions \cite{muller_toughening_2023,muller_tuning_2025}. Similarly, the addition of angle-dependent non-central constraints in simulations has been shown to alter the topology of local elastic structures and the system's bulk elasticity \cite{bantawa_microscopic_2021, van_der_meer_attraction-enhanced_2025}. These results establish the sensitivity of gel structure and mechanics to the geometry of interparticle interactions. However, controlled introduction of non-central interparticle potentials in colloidal systems often involves tedious and complex synthesis procedures. 

Here, we demonstrate a facile route to dial in non-central attractive depletion potentials in a suspension of colloidal spheres through a simple reduction in the electro-steric surface brush density of particles. This simple procedure enables activation of non-central constraints without introducing shape anisotropy, patchiness, or surface roughness. We hypothesize that potential interpenetration of brush layers between neighboring particles introduces reversible angular constraints, fundamentally altering the degrees of freedom available during aggregation and arrest. As we show, this emergent non-central interaction suppresses local densification, stabilizes low-coordination networks, and reshapes gel morphology across length scales, thereby altering the bulk elasticity. Importantly, the colloidal system employed here enables detailed three-dimensional quantification of microstructure at the single-particle level, allowing direct measurement of bond-level statistics and quantitative comparison with simulations. Leveraging this capability, we systematically validate the effect of brush-mediated non-central forces using a combination of theory, experiments, and simulations. Together, these results demonstrate how changes in brush density can introduce hierarchical structural changes throughout a colloidal depletion gel network, with direct impacts on its bulk rheology.

\section{Brush-mediated bending rigidity} 

In many classes of colloidal suspensions, a charged polymer layer is grafted onto particle surfaces to suppress aggregation and impart electro-steric stability to the suspension. We hypothesize that lowering the grafted polymer density allows for finite interpenetration between the brush layers of neighboring particles, which in turn allows for polymer-level brush–brush interactions (e.g., weak entanglements) to form between colloids. Though these interactions may be subtle at the molecular scale, they collectively introduce a constraint to rotation of contacting particles and hence result in an effective bending rigidity between them. This reduced degree of freedom at the particle level will in turn change the meso- and macro-structure of the colloidal network, establishing a direct physical pathway for tuning gel structure without invoking shape anisotropy, patchy contacts, or engineered surface topology. The overall view of particle-level architecture, and its resulting meso- and macro-scale morphologies are shown in Fig.\ref{fig:summary}. To alter the brush density, we reduce both (i) the concentration of the surface inimer that sets the number of brush grafting sites and (ii) the total monomer feed used for surface brush growth by 25\% (Fig.\ref{fig:summary}a, left; with synthesis details provided in the Methods section). Throughout this work, high brush density colloids are colored green and low-brush colloids are colored purple for visual continuity across theoretical, computational, and experimental datasets. Importantly, this method produces high-brush and low-brush particles that are geometrically indistinguishable, as confirmed by SEM (Fig.\ref{fig:summary}a, right). Remarkably, when subject to a conventional depletion framework where the attraction is controlled by overlap volume and the ensuing osmotic pressure imbalance, this subtle change in brush density results in the emergence of bending rigidity (schematically shown in Fig.\ref{fig:summary}b,left). Non-central particle-level interactions in turn give rise to ramified structures at the mesoscale for the system with reduced surface brush density, as opposed to coarser colloidal domains for the system with higher brush densities, clearly observable in confocal images in Fig.\ref{fig:summary}b, right. The 3D network reconstructions from experimental z-stacks further confirm the contrast in percolating structures between the two systems (Fig.\ref{fig:summary}c).

\begin{figure}
\centering
\includegraphics[width=1\linewidth]{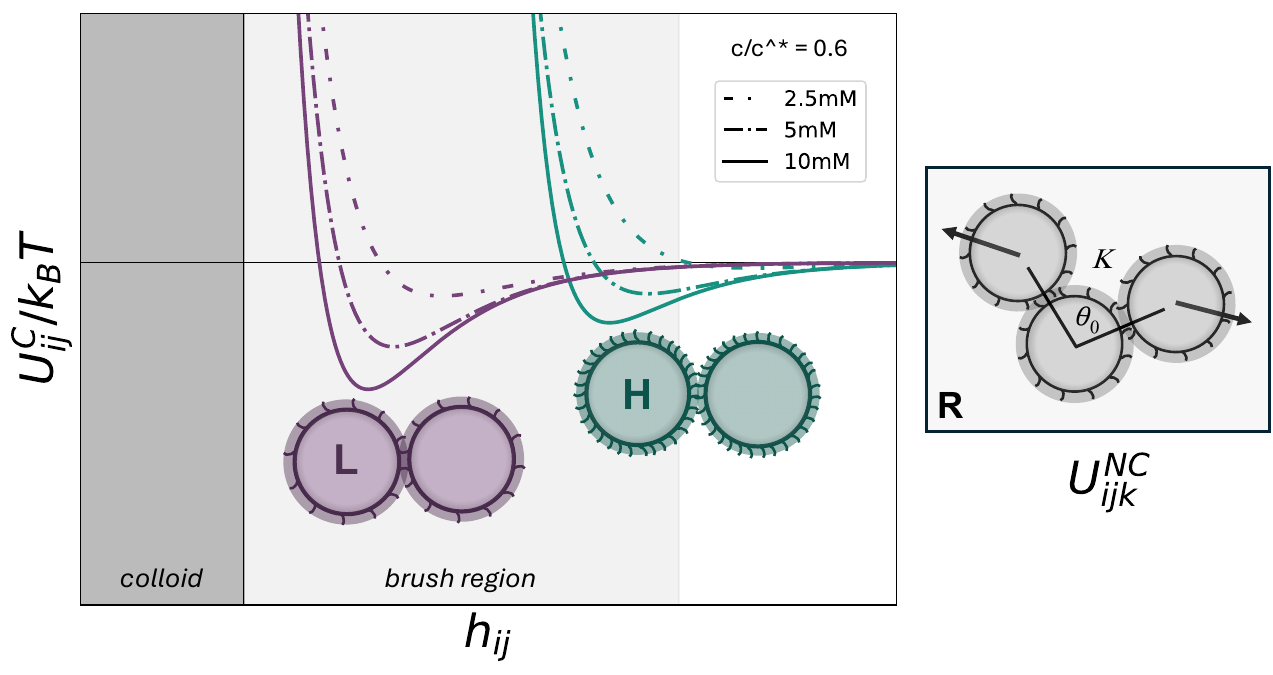}
\caption{\textbf{Brush-mediated interaction model coupling central force attraction and non-central, distance-activated bending rigidity.} Schematic interaction potentials for high-brush (H, green) and low-brush (L, purple) colloids under the experimental condition of $c/c^* = 0.6$ and 10 mM NaCl. The brush layer (light gray) defines a finite overlap region where brush–brush interactions can activate non-central constraints. Dashed curves indicate the effect of changing electrostatic screening with salt concentration. In this model, low-brush particles carry less charge and exhibit stronger net attraction under identical conditions. In the overlap regime, a non-central angular bending rigidity (R) is applied by tracking three-body angles $\theta_0$ and penalizing deviations from this configuration with a bending stiffness $K$.}
\label{fig:schematic}
\end{figure}

To test our hypothesis that rigidity emerges as a direct consequence of lowering the grafted polymer density, we develop a theoretical description of particle-level brush–brush interactions and simulate the resulting structures for large numbers of colloids. A minimal interaction model is introduced that combines conventional depletion-driven central attractions with a distance-acti\-vated angular rigidity. The physical basis of this framework, summarized schematically in Fig. \ref{fig:schematic}, is that reducing the surface brush density lowers the electro-steric repulsion between particles, allowing them to approach more closely, which in turn leads to increased interpenetration of neighboring brush layers. Consequently, the effective interparticle potential is modified through chan\-ges in the depth and position of the attractive potential well, while increased brush-brush overlap promotes the activation of angular constraints that give rise to non-central interactions.

Full details of our interaction framework are described in Supplementary Information S1 and briefly summarized here. We model the brush coating as a uniform shell of height $h_b$ surrounding each colloid. This brush region is assumed to be isotropic, consistent with the chemistry of the surface-grafted polymer layer and the absence of designed patchiness or directional motifs. Colloids interact through a short-range, pairwise central-force potential $\mathbf{U}_{ij}^{C}$ that captures the combined effect of a depletion attraction \cite{asakura_interaction_1958,vrij_polymers_1976} and a screened double-layer electrostatic repulsion \cite{derjaguin_theory_1993}. The parameters for these potentials are calculated from corresponding experimental values. A natural outcome of this model is that high- and low-brush particles experience distinct interaction landscapes, even under identical depletant and salt concentrations. Low brush density particles carry less surface charge (and thus exhibit more effective screening), leading to a higher net attraction across salt concentrations as demonstrated by the purple curves in Fig.\ref{fig:schematic}. We also introduce an electro-steric shift parameter to represent the hypothesized interpenetration of neighboring brush layers. A larger shift value moves the potential minimum towards the edge of the brush region, corresponding to minimal interpenetration at high brush densities (green, Fig.\ref{fig:schematic}). Conversely, a smaller shift leaves the potential minimum deeper inside the brush region close to the colloid surface (purple), consistent with our theory of increased brush–brush interpenetration at reduced brush densities.

The central potential is supplemented by an explicitly non-central three-body angular bending rigidity $\mathbf{U}_{ijk}^{NC}$ that only activates when particle interactions fall within the brush region. This interaction is diagrammed in Fig.\ref{fig:schematic}R. When three or more colloids form a cluster with overlapping brush regions, we record the instantaneous three-body angle $\theta_0$. This approach does not impose fixed values of $\theta_0$, allowing a distribution of bond angles to emerge naturally as particles interact. This geometric constraint is also reversible: at subsequent times $t$, if any of the pairwise bonds break, the stored angle is forgotten and rigidity is removed. If these bonds persist, any attempted deviation $\delta \theta = \theta_t - \theta_0$ from this stored geometry will result in an energetic penalty:
\begin{equation}
    \mathbf{U}_{ijk}^{NC} (K,\theta) = \frac{1}{2} K (\delta \theta)^2
        \label{eq:angular-rigidity}
\end{equation}

\begin{figure*}[ht!]
\centering
\includegraphics[width=1\linewidth]{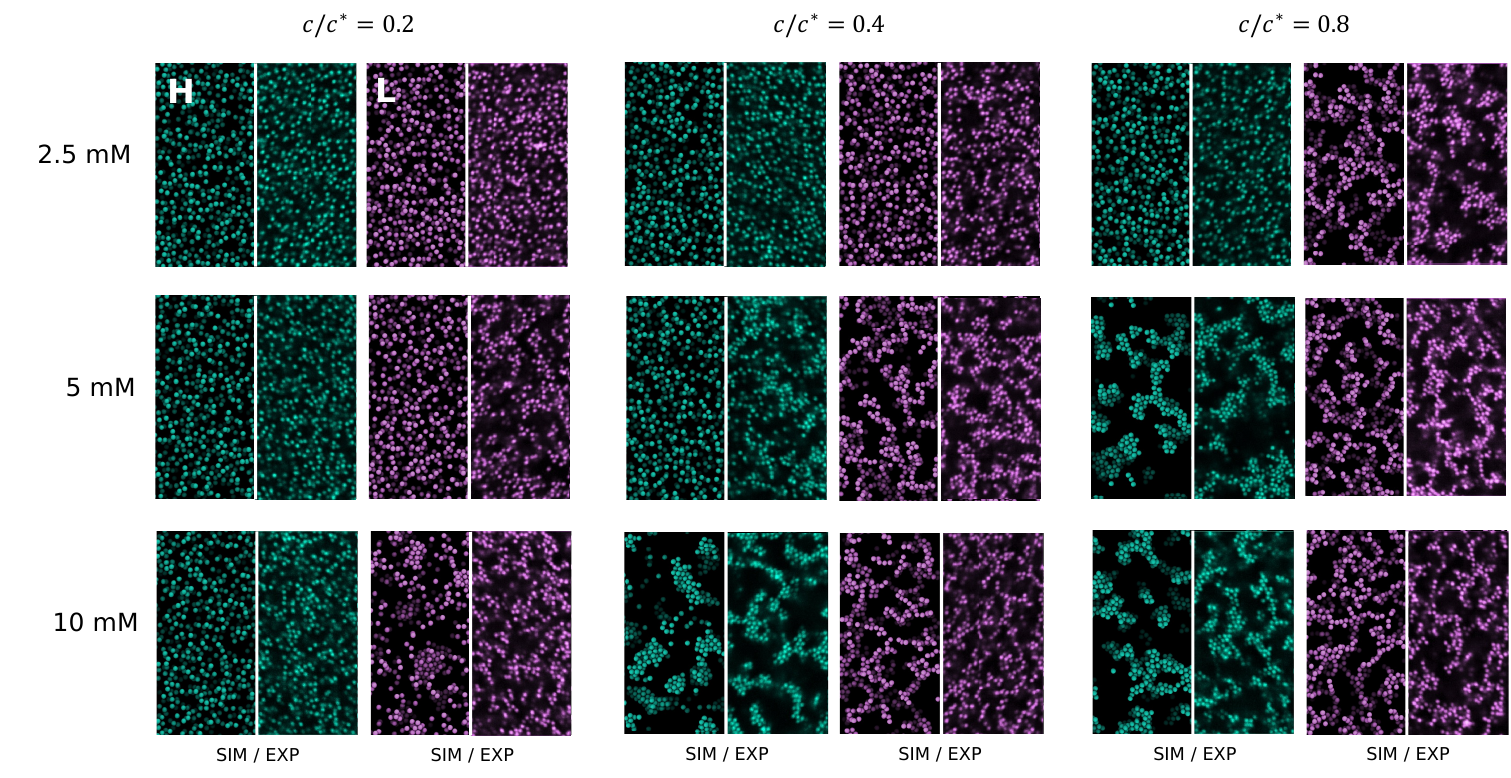}
\caption{\textbf{Brush density reshapes gel morphology across depletant-salt phase space.} Side-by-side comparison of simulated structures (left in each pair) and confocal images (right) for high-brush (green) and low-brush (purple) colloids. Rows correspond to increasing depletion strength via PAM concentration ($c/c^* = 0.2, 0.4, 0.8$), and columns correspond to increasing electrostatic screening via NaCl concentration (2.5, 5, 10 mM). Low brush density promotes earlier percolation and the formation of consistently more stringy, ramified networks. Under the same conditions, high-brush samples form dense, cluster-rich microstructures. Additional mapping is provided in Supplementary Fig.\ref{fig:SI_total_mapping}.}
\label{fig:total_mapping}
\end{figure*}

We scale the bending stiffness $K$ of this harmonic potential to be proportional to the total central attraction strength, $K=\mathbf{U}_{0}^{C}/kT$. This assumption is built on the notion that stronger central attractions between colloids are expected to increase resistance to bond bending, making angular deformation more energetically costly. For two colloids separated by a surface-to-surface distance $h_{ij}$, the total interaction potential is therefore written as
\begin{equation}
    \mathbf{U}_{ij}^{total} (h_{ij}) = 
    \begin{cases}
    \mathbf{U}_{ij}^{C}(h_{ij}) &\text{for} \enspace h_{ij} > 2h_b, \\
    \mathbf{U}_{ij}^{C}(h_{ij})+\mathbf{U}_{ijk}^{NC} &\text{for} \enspace h_{ij} \leq 2h_b
    \end{cases} 
    \label{eq:total_potential}
\end{equation}

Fig.\ref{fig:schematic} shows how these combined effects confine high-brush interactions to the edge of the brush region (green), even as salt screening increases the attraction strength. This represents limited brush-brush interpenetration, resulting in minimal non-central effects. In contrast, low-brush interactions extend substantially further through the brush region (purple), corresponding to deeper interpenetration of low-density brushes and more effective bending constraints.

\section{Brush density alters multiscale structure}  

To test our theoretical framework, a wide range of salt and depletant concentrations are experimentally and computationally tested for both brush densities, holding the particle volume fraction fixed at $\phi=0.1$. Representative snapshots from confocal imaging and simulation are shown side-by-side for nine (9) different conditions in Fig.\ref{fig:total_mapping}. A comprehensive mapping of 28 studied conditions is also provided in Fig.\ref{fig:SI_total_mapping}. For both brush densities, varying the concentrations of depletant and salt controls the central-force interaction landscape. Increasing the depletant concentration deepens the depletion attraction well, while increasing the salt concentration reduces the electrostatic barrier for aggregation. As expected, strengthening attraction through either route drives both systems from dispersed states to percolated networks. This is particularly clear in the high-brush system at intermediate depletant concentration ($c/c^* = 0.4$), where increasing the salt concentration cleanly tunes the system through three distinct regimes\textemdash{}a stable suspension, disconnected clusters, and finally a system-span\-ning gel network.

These salt and depletant trends are consistent with both systems remaining primarily governed by classical depletion-driven aggregation; however, the mapping reveals a stark difference in the arrested network morphologies between the two brush densities. At high brush density, gelled states exhibit coarse domains with locally dense clusters. These microstructures align with the established phenomenology of depletion gels at comparable volume fractions \cite{landrum_delayed_2016,mangal_small_2024,rajaram_microstructural_2010}. In striking contrast, gels made of low-brush particles consistently form networks that are more ramified, with thinner strands and visibly reduced local densification. These differences are most pronounced under conditions where attractions are strongest (high salt, and high depletant), indicating that this morphology is a robust arrested state that persists deep into the gelled regime, rather than a boundary effect. These observations are consistent with the emergence of strong non-central forces \cite{bantawa_hidden_2023,bonacci_contact_2020,bantawa_microscopic_2021}.

\begin{figure*}[t!]
\centering
\includegraphics[width=1\linewidth]{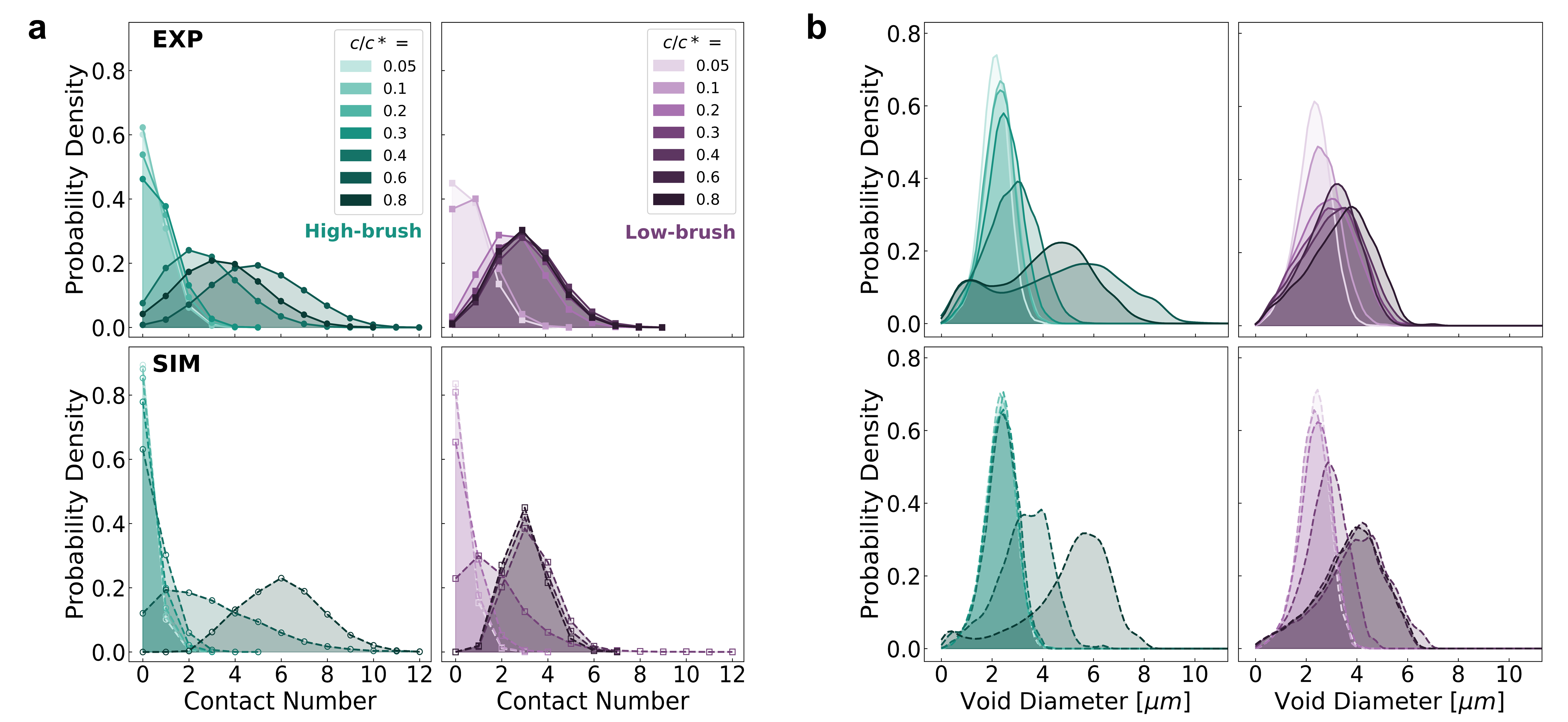}
\caption{\textbf{Brush density controls local coordination and void size in arrested gels.} Experimental (top, solid) and simulated (bottom, dashed) distributions comparing high-brush (green) and low-brush (purple) systems across PAM concentrations at 5 mM NaCl. (a) Contact number distributions shift to higher coordination with increasing attraction; high-brush gels (left) develop broader high-$Z$ tails, whereas low-brush gels (right) remain centered near $Z\approx3$. (b) Void diameter distributions broaden substantially for high-brush gels (left), indicating increased mesoscale heterogeneity, while low-brush gels (right) maintain smaller and more uniform void sizes. Data across additional salt concentrations are provided in Supplementary Figs.\ref{fig:contact_all},\ref{fig:voidsize_exp}.}
\label{fig:distributions}
\end{figure*}

We also see that reducing the brush density moves the location of the gelation boundary. Low-brush particles appear to gel under conditions where the high-brush system remains stable. For example, at 2.5 mM NaCl, $c/c^* = 0.8$ (top right), high-brush particles remain largely dispersed while low-brush particles form a percolated network.

We quantitatively validate these observations by measuring local particle volume fraction as the microscopic measure of the structure, and void size as a mesoscopic measure of the particulate network. Fig.\ref{fig:distributions} summarizes how these micro- and meso-scale structures evolve with increasing depletion strength under a fixed, intermediate salt concentration of $5 mM NaCl$. Corresponding results across all other salt concentrations are shown in Supplementary Information S7 and display similar trends. For both brush densities, we again see the consistent effect of increasing total attraction. The coordination number (Fig.\ref{fig:distributions}a) and void size (Fig.\ref{fig:distributions}b) both increase with stronger depletion interactions, reflecting the visually observed, system-wide transition from dispersed suspensions to space-spanning networks. Although both systems are clearly depletion-driven, the structural details of their networks are distinct. At the microscale, high-brush gels develop broad contact number distributions with pronounced tails toward higher coordination with $\langle Z \rangle \sim 6$. These features are consistent with the locally dense domains observed in Fig.\ref{fig:total_mapping}. In contrast, low-brush gels maintain narrow distributions centered near $\langle Z \rangle \sim 3$ (Fig.\ref{fig:distributions}a), indicating that the network is indeed dominated by loosely coordinated, strand-like motifs rather than compact local packing. This suppression of high-coordination environments provides direct quantitative evidence that the reduced brush density limits the densification pathway that normally accompanies attraction-driven aggregation in depletion systems. The stability of these networks at low average coordination across increasing attraction strengths further supports the emergence of effective non-central constraints.

This distinction is also demonstrated at the mesoscale (Fig.\ref{fig:distributions}b). High-brush gels exhibit a broad distribution of void sizes with a long tail extending to voids exceeding five particle diameters. This distribution is consistent with a heterogeneous network composed of dense domains separated by large pores. Low-brush gels, by contrast, maintain smaller and more narrowly distributed voids, consistent with the distribution of a more ramified, less spatially heterogeneous network throughout the sample volume. While the simulated structures are slightly coarser than experiments overall, the relative void-size trends remain consistent for both brush densities. This systematic offset is likely the result of the simplified hydrodynamics in our theoretical framework. Together, these results demonstrate how changes in brush-mediated bending rigidity effectively reshape local packing and mesoscale network architecture. 

\section{Changes in depletion-driven assembly}

\begin{figure*}[t!]
\centering
\includegraphics[width=1\linewidth]{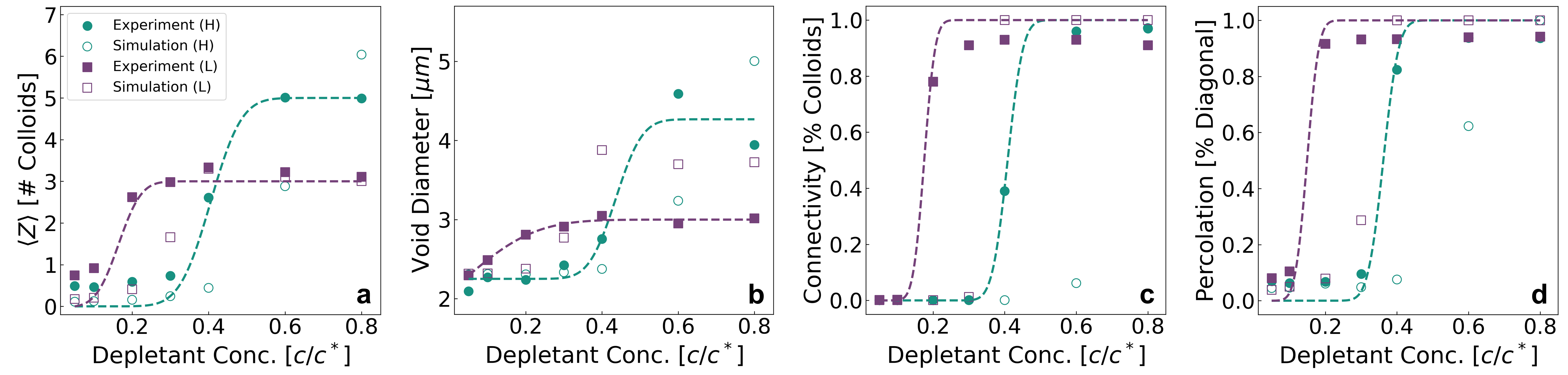}
\caption{\textbf{Multiscale structural metrics reveal earlier percolation and distinct arrested states at low brush density.} Structural metrics for high-brush (green) and low-brush (purple) systems across depletant concentrations at 5 mM NaCl, showing experiment (filled symbol) and simulated (open symbol) data. (a) Average contact number $\langle Z \rangle$ rises sharply in the high-brush system but plateaus more gradually for the low-brush case. (b) Mean void diameter increases strongly in the high-brush system while remaining comparatively uniform at low brush density. (c) Fraction of particles in the largest connected component (LCC) and (d) percolation fraction, largest-cluster diameter normalized by the sample box diagonal, both rise sharply at lower depletant concentration in the low-brush system, consistent with the earlier formation of a system-spanning network. Dashed curves are provided as visual guides to trends in the experimental data. Additional salt concentrations are provided in Supplementary Fig\ref{fig:full_data}.}
\label{fig:Quantitative Data}
\end{figure*}

Having examined the distributions of contact number and void size as local and mesoscopic measures of colloidal gel structure, we next examine the macroscpic features of the system. This is quantified in addition to average contact numbers and void sizes, through calculation of the lengthscale and the fraction of colloids that belong to the largest connected component in the system, shown in Fig.\ref{fig:Quantitative Data}, with full mapping provided in Fig.\ref{fig:full_data}). The fraction of total particles contained in the largest connected component (LCC) reveals whether colloids form a connected network rather than remaining in disconnected clusters, while the size of LCC (percolation), normalized by the sample box diagonal, provides a direct measure of whether this component spans the entire sample. Both measures approach 100\% with increasing depletant concentration. This indicates the formation of depletion-driven space-spanning networks that incorporate nearly all observable particles. Notably, the complete percolation of both systems shown in Fig.\ref{fig:Quantitative Data}c,d indicates that the observed structural differences between the high- and low-brush systems emerge within fully percolated networks. These distinctions persist in dynamically arrested states, as demonstrated by videos of simulated high- and low-brush particle trajectories, described in more detail in Supplementary Information S5. This interpretation is also supported experimentally by dynamical measurements (Fig.\ref{fig:full_data}.e), where the MSD slope falls below unity after percolation in both systems, consistent with sub-diffusive motion in arrested gels. Comparing connectivity with average contact number $\langle Z \rangle$ (Fig.\ref{fig:Quantitative Data}a) and mean void diameter (Fig.\ref{fig:Quantitative Data}b) reveals a consistent multiscale picture: the structure of the two different brush systems only diverges significantly after percolation.

Another key result is that low-brush networks percolate at lower depletant concentrations than high-brush networks. This indicates an earlier onset of gelation under comparable experimental conditions, consistent with the combined effects of reduced electrostatic repulsion, which lowers the aggregation threshold, and enhanced bending rigidity, which promotes more efficient network formation. Experimentally, the high-brush particle structure exhibits a delayed rise in $\langle Z \rangle$, remaining low until $c/c^* \geq 0.4$ before increasing sharply and saturating near $\langle Z \rangle \approx5$. This increase aligns with the percolation trend observed in the connectivity metrics. In contrast, the low-brush system shows a more gradual increase in $\langle Z \rangle$ toward a plateau near $\langle Z \rangle \approx3$, with the transition shifted to lower depletant values. 

\begin{figure*}
\centering
\includegraphics[width=1\linewidth]{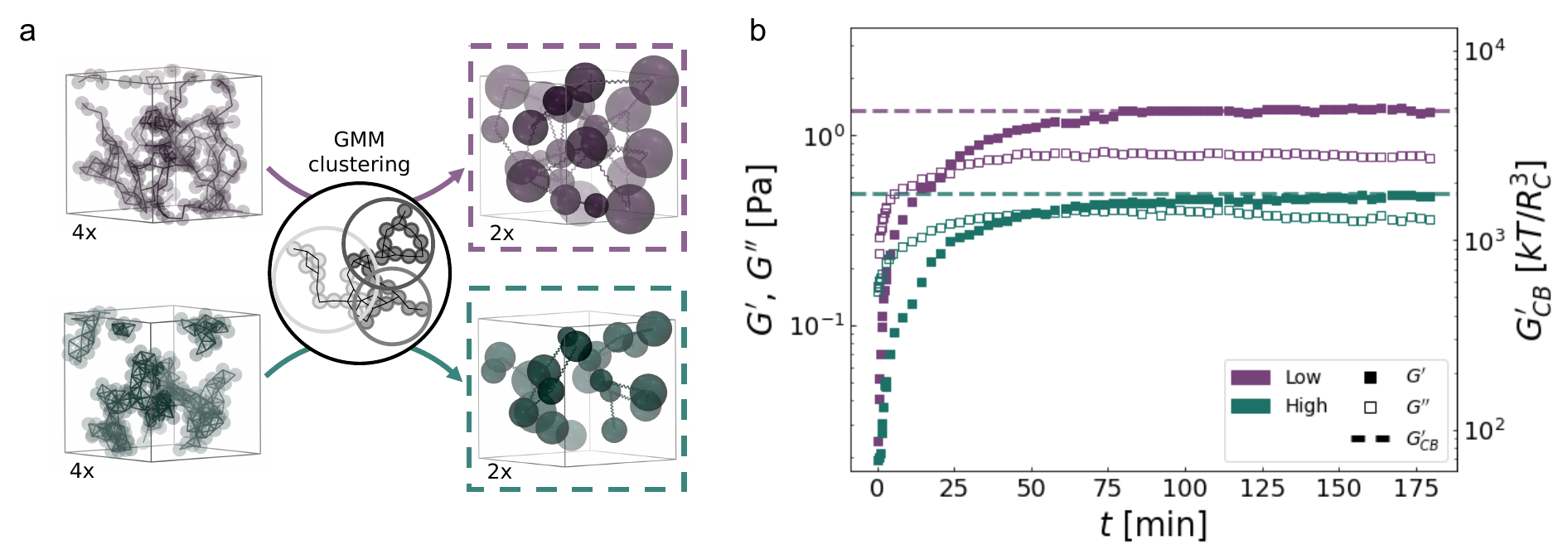}
\caption{\textbf{Brush density tunes gel elasticity through mesoscale network structure.} (a) Schematic illustration of clustering method from individual particles into mesoscale cluster networks. The particle-resolved low-brush (purple) and high-brush (green) networks are grouped into locally rigid domains using a Gaussian Mixture Model (GMM). These clusters are then represented as overlapping spheres in a 3D cluster-level network (in the 3D diagram, sphere volume has been reduced for visual clarity). This network is used to calculate the modulus from Cauchy-Born theory. (b) Comparison of theoretical predictions with time-sweep SAOS at $\phi=0.1$, 10 mM NaCl, and $c/c^* = 0.6$ for both low-brush and high-brush gels. Filled squares and open squares show experimental \(G'\) and \(G''\), respectively (left axis). The low-brush gel exhibits an earlier \(G' = G''\) crossover and a higher long-time plateau modulus than the high-brush gel. Dashed lines show the matching cluster-network predictions of the storage modulus, $G'_{CB}$, obtained from simulated structures (right axis).}
\label{fig:rheology}
\end{figure*}

Importantly, our theoretical framework correctly predicts this result. Reduced brush densities lower both steric and electrostatic repulsions, yielding stronger net attraction. At $c/c^* = 0.2$ and 5 mM NaCl, the model predicts $U_0^C\sim 2 kT$ for high-brush and $U_0^C\sim 5 kT$ for low-brush particles, reflecting the observed change in gelation boundary. In fact, this shift persists in simulations with and without bending rigidity (Fig.\ref{fig:SI_norigidity}), clearly demonstrating that the observed percolation of low-brush systems at lower polymer concentrations is not a feature of non-central forces, but because of the brush layer's sensitivity to charge screening. Near the gel boundary however, both systems exhibit a regime in which networks become nearly fully connected before $\langle Z \rangle$ saturates. This indicates the presence of large, loosely coordinated clusters that approach percolation at relatively weak attractions ($ U_0^C \sim  1$-$7 kT$). This result is consistent with the state diagram for depletion-driven colloidal gels \cite{lu__colloidal_2013}. Simulations reproduce these trends with expected discrepancies at the weakest attractions, where complete hydrodynamic interactions are known to play a larger role in gelation \cite{varga_hydrodynamics_2015}. Nevertheless, across the experimentally relevant regime, the agreement between experiments and simulations demonstrates that a minimal bending-rigidity framework is sufficient to reproduce the coupled evolution of microstructure, mesoscale heterogeneity, and macroscale connectivity in this system.

\section{Brush-mediated rheological behavior} 

Finally, having demonstrated the structural contrasts for the two brush densities, we show that invoking the bending rigidity also serves as a direct and predictive control parameter for bulk viscoelasticity of colloidal gels. It is well-established that the mechanics of soft particulate gels are governed by their underlying network structure, and that deformation in these systems is largely carried by a hierarchical backbone rather than individual contacts alone \cite{nabizadeh_network_2024, rouwhorst_nonequilibrium_2020,zaccone_elasticity_2009,whitaker_colloidal_2019,bantawa_hidden_2023,rocklin_elasticity_2021}. Here we estimate this backbone using a clustering process illustrated schematically in Fig.\ref{fig:rheology}a. Particles are grouped into locally rigid clusters using a network-based Gaussian Mixture Model (GMM) \cite{nabizadeh_network_2024}. These clusters are then assembled into a new network of representative spheres connected by inter-cluster bonds. Visualizing this network clearly demonstrates that the low-brush system contains a more densely connected cluster-level network than the corresponding high-brush system. Although the thin strands of the low-brush particle network appear more tenuous than the locally packed clusters of the high-brush system, we find that the narrower void distribution and lower mean void size seen in Fig.\ref{fig:distributions}b and Fig.\ref{fig:Quantitative Data}b effectively produce a denser mesoscale backbone. We also validate that the introduction of non-central constraints mechanically stabilizes the low-brush clusters at low internal coordination number (details provided in Methods). Once clusters are identified as the main source of rigidity in the global network, the storage modulus $G'$ can be estimated using the Cauchy–Born free energy expansion for metastable amorphous solids, including both central $(G'_C)$ and bending $(G'_B)$ contributions \cite{zaccone_shear_2009, zaccone_elasticity_2009} as
\begin{equation}
    G'= G'_C+G'_B= \left( \frac{4}{5\pi}\kappa_s+\frac{124}{135\pi}\kappa_b \right) \frac{\langle Z^c \rangle \phi^c}{\langle \xi^c \rangle}
        \label{eq:cauchy-born}
\end{equation}
Here $\langle Z^c \rangle$ is the average coordination number between clusters, $\phi^c$ is the cluster volume fraction, and $\langle \xi^c \rangle$ is the average cluster radius. Since inter-cluster bonds occur between component particles, the bond ``stretch" stiffness $\kappa_s$ is estimated from equipartition using the pairwise separation distances measured in the particle network: $\kappa_s\approx kT/(\langle r_{ij}\rangle^2 - \langle r_{ij}^2 \rangle)$. Similarly, the ``bend" stiffness is estimated from the bending rigidity parameter $K$ normalized by the variance of separations between angular neighbors in each i-j-k particle triplet: $\kappa_b\approx K/(\langle r_{jk}\rangle^2 - \langle r_{jk}^2 \rangle)$, where $i$ is the central particle and $r_{jk} = \sqrt{(r_{ij})^2+(r_{ik})^2-2r_{ij}r_{ik}cos\theta}$. 

Fig.\ref{fig:rheology}b shows a comparison of this mean-field prediction with the viscoelastic moduli measured during a small amplitude oscillatory shear (SAOS) time-sweep experiment under identical preparation conditions ($\phi=0.1$, 10 mM NaCl, $c/c^* = 0.6$). While the moduli calculated from the clustering and Cauchy-Born theory are in simulation units (shown using horizontal dashed lines in Fig.\ref{fig:rheology}b), they precisely track the role of non-central bending forces. Experimentally, both high-brush and low-brush gels initially exhibit predominantly viscous behavior (\(G''>G'\)), followed by a progressive rise in elasticity as the network forms. The low-brush system exhibits accelerated evolution consistent with its more ramified aggregation pathway arising from enhanced bending rigidity: it reaches the \(G' = G''\) crossover at $\approx$25 min, compared to $\approx$60 min for the high-brush system, indicating faster growth of the effective network volume fraction and earlier percolation. Additionally, at long times, the low-brush gel plateaus at a significantly larger storage modulus, fully consistent with our structure-based mean-field predictions. Specifically, simulations predict $G'_L\approx2.7G'_H$, compared to the experimentally measured increase of $G'_L\approx2.8G'_H$. This agreement indicates that our theoretical model correctly identifies the dominant mechanisms that govern stiffness enhancement from brush-density reduction.

Decomposing the theoretical contributions, we find that the change in interaction landscape not only accelerates gelation, but also directly impacts the total modulus by changing the bond stiffness between clusters. The increase in total attraction results in a larger effective $K$, thus increasing the bending stiffness $\kappa_b$. Simultaneously, stronger central-force interactions also increase the estimated stretch stiffness $\kappa_s$ via equipartition: a deeper and narrower attractive well (Fig.\ref{fig:schematic}) decreases the variance in particle separations, indicating a higher $\kappa_s$. Together, these bond-level changes contribute to an increase in $G'$ when brush density is decreased.

The remaining stiffness enhancement originates at the mesoscale. Compared to the high-brush system, the low-brush gel forms approximately $\sim$1.2× more clusters, leading to a $\sim$1.6× increase in the effective cluster volume fraction $\phi^c$. Although the more ramified network also exhibits a lower mean cluster–cluster coordination, it nevertheless contains nearly twice as many unique inter-cluster connections. Increasing the distribution of these inter-cluster connections across the network produces a more uniform and redundant backbone, which can further enhance the macroscopic elastic response.

Taken together, our results clearly demonstrate that non-central constraints to particle motion can be invoked in simple depletion gels through the manipulation of polymer brushes grafted on the surface of colloids. We demonstrate that this behavior occurs even when particles are outwardly similar in shape and size, revealing a surface-mediated interaction that standard, centrosymmetric depletion models miss. By allowing for the individual polymer chains to interpenetrate the brush layer of a neighboring colloid and their interactions, an effective bending rigidity emerges between neighboring colloids. Additionally, lowering the polymer brush density also slightly changes the extent of attraction between colloids, and is shown to directly control the structure of the particulate system at all length scales. At the microscale, this bending rigidity results in suppressed number of contacts, with limited number of bonded neighbors ($\langle Z \rangle \sim 3$), as opposed to higher coordination numbers of $\langle Z \rangle \sim 6$ for the higher brush densities. At the mesoscale, these lower number of contacts and mobility-constraints result in highly ramified structures that are reminiscent of protein gels \cite{ako_salt-induced_2010,rouwhorst_protein_2021}, shown through narrow distributions of void sizes with smaller averages compared to coarser compact domains of high-brush colloids, typical for depletion gels. The ramified stringy networks of low-brush colloids form space-spanning percolated structures more effectively, and at lower concentration of depletant as well. As a result, the cluster-level network for the low brush density particles [with enahnced bending rigidity] construct a highly connected rigid network of signiticantly higher storage modulus compared to higher brush density particles with minimal bond bending constraints. This demonstrates a practical framework for systematic tuning of particle-level interactions, manifesting in strikingly different morphologies and mechanics/rheologies at the meso- and macro-scales. More broadly, this work demonstrates that interaction symmetry - not only interaction strength - can be continuously tuned in otherwise isotropic particulate systems. By introducing angular constraints without altering particle geometry, we access a regime of emergent non-central forces that reorganize rigidity, connectivity, and elasticity across scales. This mechanism provides a new physical pathway for programming mechanical response in disordered matter, bridging colloidal gels, biopolymer networks, and amorphous solids. The ability to decouple force symmetry from particle shape in depletion systems opens new opportunities for designing soft materials whose mechanical properties are encoded directly at the level of interparticle constraints.

\section{Methods}
\phantomsection
\subsection{Core-shell particle synthesis}
Monodisperse colloidal particles of diameter 1.06 µm (dispersity = 0.03) were synthesized following a modified protocol from Kodger et al. \cite{kodger_precise_2015} and Park et al. \cite{park_aqueous_2018}. The process involved three primary steps: (1) synthesis of fluorescent core particles, (2) growth of non-fluorescent shells, and (3) surface functionalization with charged stabilizers. All chemicals were purchased from Sigma Aldrich unless otherwise specified.

\subparagraph{\textbf{Core synthesis:}}
Fluorescent cores were synthesized using a 45:55 volumetric ratio of 2,2,2-trifluoroethyl meth\-acry\-late (TFEMA, 3.21 mL) and tert-butyl methacrylate (tBMA, 3.93 mL), along with ethylene glycol dimethacrylate (EGDMA, 0.147 mL) as a crosslinker. Fluorescence was incorporated by adding coumarin-methacrylate (2.5 mL, synthesized according to Kodger et al. \cite{kodger_precise_2015}). Initiation was carried out using 2,2'-azobis(2-methylpropio\-nitrile) (AIBN, 0.077 g), with 3-sulfopropyl methacrylate potassium salt (0.077 g) and 2-(2-bromoisobutyryloxy) ethyl acrylate (0.143 mL, synthesized according to Kodger et al. \cite{kodger_precise_2015}) added for surface functionality. All reagents were dispersed in a methanol–water mixture (170.2 mL methanol, 37.1 mL water), and polymerization proceeded at 80 $^\circ C$ under reflux for 5 hours with continuous stirring.

\subparagraph{\textbf{Shell grown:}}
A non-fluorescent shell was grown on the core using the same TFEMA (2.70 mL)–tBMA (3.30 mL) ratio with poly(vinylpyrrolidone) (PVP, 2.44 g) as a steric stabilizer. The core suspension (15 mL, 20 vol\% in methanol-water) and 2-(2-bromoisobutyryloxy) ethyl acrylate (0.36 mL) were added to a methanol (155.4 mL)–water (11.9 mL) mixture. Polymerization was initiated with AIBN (0.076 g) at 55 $^\circ C$ for 16 hours.

\subparagraph{\textbf{Surface functionalization:}}
Charged stabilizers were grafted onto the particles using a 1:1 molar ratio of 2-acrylamido-2-methyl-1-propanesulfonic acid sodium salt (SPAm, 2.52 mL) and N,N-dimethylacrylamide (DMA, 0.678 mL), polymerized in the presence of copper(I) chloride (0.028 g), copper(II) chloride (0.035 g), and hexamethyltriethylenetetramine (0.151 mL). The core-shell suspension (15 mL, 25 vol\% in methanol-water) and a sacrificial initiator, PEGini (0.289 mL, synthesized according to Kodger et al. \cite{kodger_precise_2015}), were added to a methanol (6.74 mL)–water (6.68 mL) mixture. The reaction proceeded for 6 hours at room temperature.

To prepare particles with reduced surface brush density, we modified both the shell-growth and surface-func\-tion\-al\-iza\-tion steps. During shell growth, the amount of 2-(2-bromo\-iso\-butyryloxy) ethyl acrylate was reduced by 25\% (from 0.36 mL to 0.27 mL) to decrease the density of brush initiation sites on the particle surface. The subsequent surface functionalization was carried out identically to the high-brush case but with a 25\% reduction in total monomer feed, using 1.89 mL of SPAm and 0.509 mL of DMA. Together, these adjustments yielded particles with consistently reduced brush density.

\subsection{Preparation of colloidal gels}
A 75/25 wt\% glycerol-water mixture was used as the solvent, providing a close refractive index and density match to the particles. Non-adsorbing polyacrylamide (PAM, Mn = 150,000, $c^*$ = 9.93 mg/mL \cite{park_aqueous_2018}) and sodium chloride (NaCl) were dissolved in the solvent to prepare stock polymer depletant and salt solutions. These stock solutions were then used in varying amounts to achieve the desired polymer ($c/c^*=0.05,0.1,0.2,0.3,0.4,0.6,0.8$) and salt (2.5, 3.5, 5, 10 mM) concentrations in the final gel samples.

To prepare each sample, stock solvent, polymer depletant, salt, and particle suspension were mixed in a vial to create a 0.25 mL final suspension with a particle volume fraction of 0.10. The suspension was thoroughly homogenized using a vortex mixer to ensure uniform dispersion. After mixing, the suspension was transferred into a customized glass chamber designed for microscopic imaging. Samples were left undisturbed for one hour to allow gelation before imaging.

\subsection{Confocal microscopy, structural measurements, and dynamical measurements}
A confocal laser scanning microscope (FLUOVIEW FV\-3000, Olympus) equipped with a 40× oil-immersion objective was used to acquire images of all samples. To capture the three-dimensional (3D) structure, a series of two-dimensional (2D) images was collected at vertical (z) intervals of 0.25 µm per step, starting 5 µm above the coverslip, covering a total depth of 30 µm. The imaging window in the x and y directions was 50 × 50 µm. For each sample, three z-stacks were obtained at different locations.

To perform structural measurements, a Python-based particle tracking algorithm (TrackPy) \cite{crocker_methods_1996, allan_soft-mattertrackpy_2025} was used to locate particle centers with subpixel accuracy. Particles were identified as nearest neighbors if their interparticle separation fell within the defined search distance criteria (see Supplementary Information S2 for details). Custom Python structural analysis algorithms were then used to evaluate the structure of the sample, including particle contacts, void sizes, and network connectivity.

To measure particle dynamics, two-dimensional (2D) time-lapse images were acquired in a fixed z-plane, 15 µm above the coverslip. For each sample, three image sequences were recorded at 1 frame per second for 60 seconds. Particle centers were identified using TrackPy and tracked over time to compute mean-squared displacements.

\subsection{Zeta potential}
The zeta potential of the colloidal particles was measured using a Nanoparticle Analyzer (nanoPartica SZ-100V2 Series). For analysis, the particles were diluted to a volume fraction of 0.001 in 9.5 mM Tris buffer (pH 7.5). Particles synthesized with higher surface brush density exhibited a zeta potential of –50 mV, whereas those with reduced brush density showed a less negative potential of –34 mV.

\subsection{Rheology}
Small-amplitude oscillatory shear (SAOS) experiments were performed using a Discovery Hybrid Rheometer (HR-3, TA Instruments) to characterize the viscoelastic response and gelation dynamics of the colloidal suspensions. A 40 mm parallel-plate geometry was used for all measurements. The colloidal sample was first placed on the stationary base, after which the gap between the upper plate and the base was set to 500 µm. Excess material around the edges was carefully trimmed, and a custom-designed solvent-trapping cap was used to prevent evaporation during long-term measurements. Prior to each measurement, all samples were pre-sheared at a shear rate of 50 $s^{-1}$ for 60 s to eliminate any pre-existing structures and ensure a reproducible initial state. Time-sweep measurements were then performed to monitor both the storage (\(G'\)) and loss (\(G''\)) moduli of the sample over 180 min at 0.5\% strain, 1 $rad/s$, and 20 $^\circ C$.

\subsection{Simulations}
All simulations were performed with the open-source molecular dynamics toolkit HOOMD-blue \cite{anderson_hoomd-blue_2020}. This software was modified to implement our theoretical framework. Monodisperse colloidal particles with radii $R_C=1$ and a total particle volume fraction $\phi=0.1$ were initialized in a cubic box of side length $L=30R_C$ with Lees-Edwards periodic boundary conditions. The system was allowed to briefly equilibrate under a Brownian thermostat $k_BT=0.1$ and short-range hard-sphere repulsion without the addition of attractive interactions.

The pairwise central force contribution $\mathbf{U}_{ij}^{C}$ is calculated as the sum of depletion attraction and electrostatic repulsion. We use experimental parameters to calculate the total interaction potential for each system and apply shift factors to account for the change in brush grafting density between the high-density and low-density systems, as detailed in Supplementary Information S1.

We delineate the polymer brush coating as a uniform shell surrounding each particle. The brush layer for all systems is defined with a height $h_b=20nm$ and angular bending rigidity is applied for any triplets of three particles interacting within this region. Particle-level, brush-brush interactions are defined with a bending rigidity proportional to the total magnitude of the central force minimum, $K=U_0^C/kT$. The final datasets were run for $160\tau_D$, where $\tau_D$ is the bare-particle Brownian diffusion time of a single colloid. Static structures were visualized with the molecular graphics program VMD \cite{humphrey_vmd_1996} and then recolored in Adobe Photoshop to distinguish the two brush densities.

\subsection{Identifying locally rigid clusters}
A network-based clustering technique \cite{nabizadeh_network_2024} is used to group particles into locally rigidified clusters. Simulated particle positions are embedded in topological space using the automated node2vec feature learning algorithm, which classifies nodes into topolgical neighborhoods. The optimal number of clusters is determined from the Bayesian Information Criterion (BIC), and particles are assigned to these clusters using a Gaussian Mixture Model (GMM), which assumes the total distribution of particle positions can be represented as a mixture of different Gaussian distributions, representing distinct clusters. Each cluster is then examined in real-space to ensure it represents a realistic, physically connected subgraph.

To calculate the storage modulus, locally rigid clusters are identified using the isostaticity criterion for adhesive systems taken from Phillips-Thorpe theory, $z_{iso}=2.4$ \cite{valadez-perez_dynamical_2013}. We validate the choice of this criterion by calculating the Maxwell-Calladine relation with bending constraints
\begin{equation}
    \frac{F-S}{N} = (d+r)-\frac{\langle Z \rangle c}{2}-m
    \begin{cases}
    >0 &\text{floppy} \\
    \geq 0 &\text{rigid} \\
    \end{cases} 
        \label{eq:maxwell-calladine}
\end{equation}
where $F$ is the number of floppy modes that can deform without requiring elastic energy or changing bond lengths, $S$ is the number of sites of self-stress where interparticle bonds are being compressed or stretched, and $N$ is the number of particles in the cluster. These values are estimated from the number of dimensions of the system $d$, the average coordination number in each cluster $\langle Z \rangle$, the number of rotational degrees of freedom $r$, the number of independent force components a particle can transmit $c$,  and the number of angular constraints $m=N_{\theta}/N$, defined as the number of angles in the cluster divided by the number of particles in the cluster. The isostatic limit occurs when $(F-S)/N=0$, and can be used to calculate the isostaticity criterion for purely angular constraints, where $r=0$, and $c=1$. We find that $z^{\theta}_{iso}=6-(2N_{\theta}/N) \leq 2.4$ for all clusters in our system, confirming that Phillips-Thorpe theory is an appropriate, conservative criterion for local rigidity in these brush-mediated angular bending rigidity systems.

\section*{Data Availability}
The data supporting the findings of this study are available from the corresponding authors upon reasonable request.

\section*{Code Availability}
The codes used in this study are available from the corresponding author upon reasonable request.

\section*{Acknowledgments}
Financial support for this study was provided to S.J. and A.M. by the National Science Foundation (PMP-2025613). Z.Z. and A.M. acknowledge the use of facilities and instrumentation at the UC Irvine Materials Research Institute (IMRI), which is supported in part by the National Science Foundation through the UC Irvine Materials Research Science and Engineering Center (DMR-2011967).


\putbib

\end{bibunit}


\clearpage
\onecolumn

\titleformat{\section}
  {\normalfont\Large\bfseries}
  {}{0pt}{}
\titleformat{\subsection}
  {\normalfont\large\bfseries}
  {}{0pt}{}
\titlespacing*{\section}{0pt}{2.5ex}{1ex}
\titlespacing*{\subsection}{0pt}{2ex}{0.8ex}

\renewcommand{\thesection}{S\arabic{section}}
\renewcommand{\thefigure}{S\arabic{figure}}
\renewcommand{\theequation}{S\arabic{equation}}

\begin{bibunit}

\section{Supplementary Information}

\vspace{0.8cm}

\tableofcontents
\counterwithin{figure}{section}
\renewcommand{\thesubsection}{S\arabic{subsection}}

\newpage
\phantomsection
\subsection{Numerical Model}\label{supp-numerical-model}

\renewcommand{\thefigure}{S\arabic{figure}}
\renewcommand{\theequation}{S\arabic{equation}}

We numerically represent particle interactions using a combination of central $\mathbf{U}_{ij}^{C}$ and non-central $\mathbf{U}_{ijk}^{NC}$ contributions:

\begin{equation}
    \mathbf{U}_{ij}^{total} (h_{ij}) = \mathbf{U}_{ij}^{C}+\mathbf{U}_{ijk}^{NC}
    \label{eq:total_potentialSI}
\end{equation}
\\
Each colloid is defined as a hard sphere of radius $R_C$, covered in a thin polymer brush coating of height $h_b$. In the simulation calculation all distances are non-dimensionalized by the experimental colloid radius $R_C=550nm$; however, they are reported in real units for clarity here. Based on the synthetic procedures outlined in the Methods section, we assume a short brush height of $h_b=20nm$ for all particles. The total central potential $\mathbf{U}_{ij}^C$ is the sum of two pair-wise potentials that depend on the interparticle separation distance $r_{ij}$: a depletion attraction $U_{ij}^D$ and an electrostatic repulsion $U_{ij}^E$. To account for the steric repulsion produced by overlapping brush regions we redefine these potentials in terms of the bare-colloid surface-to-surface separation distance $h_{ij} = r_{ij} - 2R_{C}$. We mimic the effect of changing brush density by moving the location of the potential minimum within the brush region, from $h_{ij}$ to a value $h_{ij}+2c_0$, where $c_0\leq h_b$. The exact value of $c_0$ emerges from independent shift factors that are applied to each component of the total central force, described below. Depletion attraction is modeled with a shifted Morse Potential:

\begin{equation}
    U_{ij}^D (h_{ij}) = D_0 k_B T (e^{-2\alpha (h_{ij}-2r_0)}- 2e^{-\alpha (h_{ij}-2r_0)})
    \label{eq:morse}
\end{equation}
\\
where $D_0$ is the attractive potential minimum, $k_BT=0.1$ is the temperature of the system, and $\alpha$ is the Morse parameter controlling the range of attraction. 

\begin{figure*}[h!]
\centering
\includegraphics[width=1\linewidth]{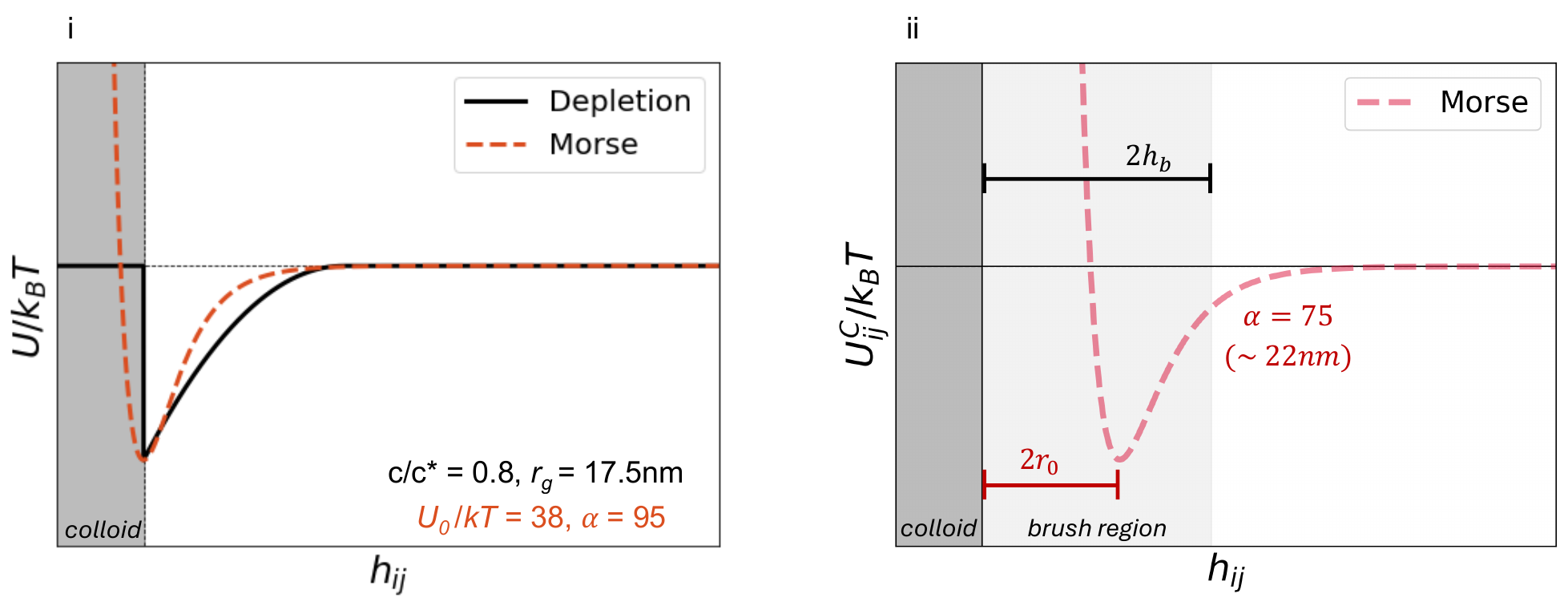}
\caption{\textbf{Numerical depletion attraction.} (i) Example scaling of the Morse Potential vs. Asakura-Oosawa Depletion Potential for $c/c^* = 0.8$. (ii) Example of the shifted Morse potential for high brush density particles, using $r_0=12nm$.}
\label{fig:SI_morse}
\end{figure*}

To correctly scale the Morse potential with experimental conditions we calculate each system's $D_0$ using the penetrable hard-sphere model from Asakura-Oosawa-Vrij theory \cite{asakura_interaction_1958,vrij_polymers_1976}. Fig.\ref{fig:SI_morse}.i shows an example of this scaling. We use the depletant overlap concentration of $c^*\approx9.93$g/mL \cite{park_aqueous_2018} to approximate the radius of gyration as $r_g=17.5nm$. This depletant size and the nominal colloid radius, $R_{nom}=R_C+h_b$, are used to calculate the osmotic pressure and the overlap volume at particle-particle contact, giving us the maximum depletion attraction $D_0$ for each depletant concentration in the experiments. We scale $\alpha$ with our depletant-to-colloid size ratio $\Delta=r_g/R_C$ in accordance with previous literature \cite{zia_micro-mechanical_2014}. We extend this range slightly to account for the $r_0$ shift described in the next paragraph, and apply the same depletion attraction range in both brush systems: $\alpha=75$, or $\sim$$22nm$.

\begin{figure*}[h!]
\centering
\includegraphics[width=1\linewidth]{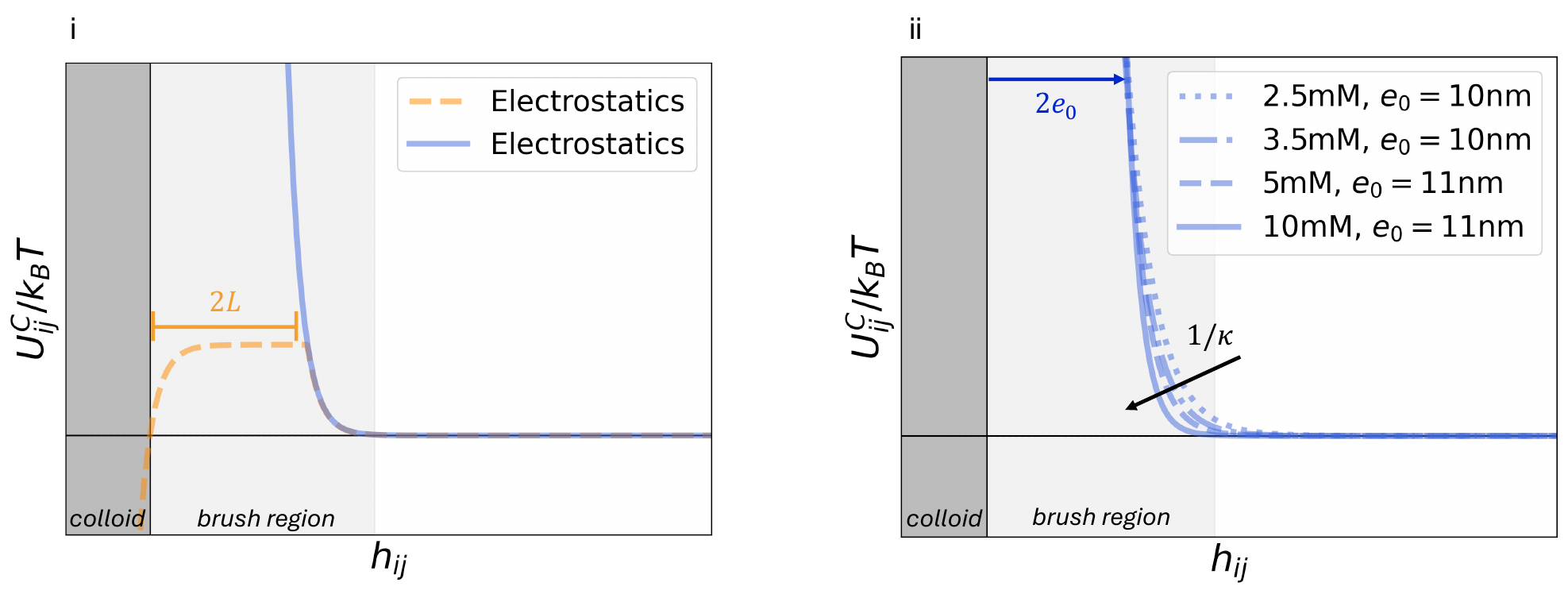}
\caption{\textbf{Numerical electrostatic double layer repulsion. }(i) Example of the charge-distributed Double Layer potential (yellow, dashed, Eq. \ref{eq:chargedistribution}) for high brush density particles with $L^H=14nm$, 
as compared with the fit of a more numerically stable shifted Double Layer potential using $e_0$=10 nm (blue, solid line, Eq. \ref{eq:doublelayer}). Both potentials are calculated for 10mM NaCl. (ii) Example of electrostatic dependence on salt concentration. Increasing salt concentration has a very small effect on the effective position of the charge surface $e_0$, and decreases the Debye length $1/\kappa$.}
\label{fig:SI_electrostatics}
\end{figure*}

To account for changes in brush density, we introduce the term $r_0$, which shifts the location of the potential minimum within the brush region. We define $r_0$ as a per-particle offset distance that depends on brush density, representing the point where brush-brush steric repulsion takes effect: $r_0^H=12nm$ and $r_0^L=4nm$. Fig.\ref{fig:SI_morse}.ii shows an example of this shift for high brush density particles. Similarly, electrostatic repulsion is modeled with a shifted Double Layer potential:
\begin{equation}
U_{ij}^E (h_{ij}) = \frac{1}{2}(R_C + e_0) \frac{Z}{k_B T} e^{-\kappa (h_{ij} - 2 e_0)} 
\label{eq:doublelayer}
\end{equation}
where $Z$ is the electrostatic interaction constant calculated from the Zeta potential using Gouy-Chapman theory, $k_B$ is the Boltzmann's constant, $T$ is the temperature, and $\kappa$ is the inverse Debye screening length calculated for a given salt concentration as:
\begin{equation}
\kappa = \frac{2zc_s N_A e^2 }{\varepsilon_r \varepsilon_0 k_B T}
    \label{eq:kappa}
\end{equation}
where $z$ is the ion valency, $c_s$ is the concentration of salt in $mol/m^3$,  $N_A$ is Avogadro's number, $e$ is the elementary charge of an electron, $\varepsilon_r$ is the dielectric constant of the solvent, and $\varepsilon$ is the permittivity of free space.

\begin{figure*}[h!]
\centering
\includegraphics[width=1\linewidth]{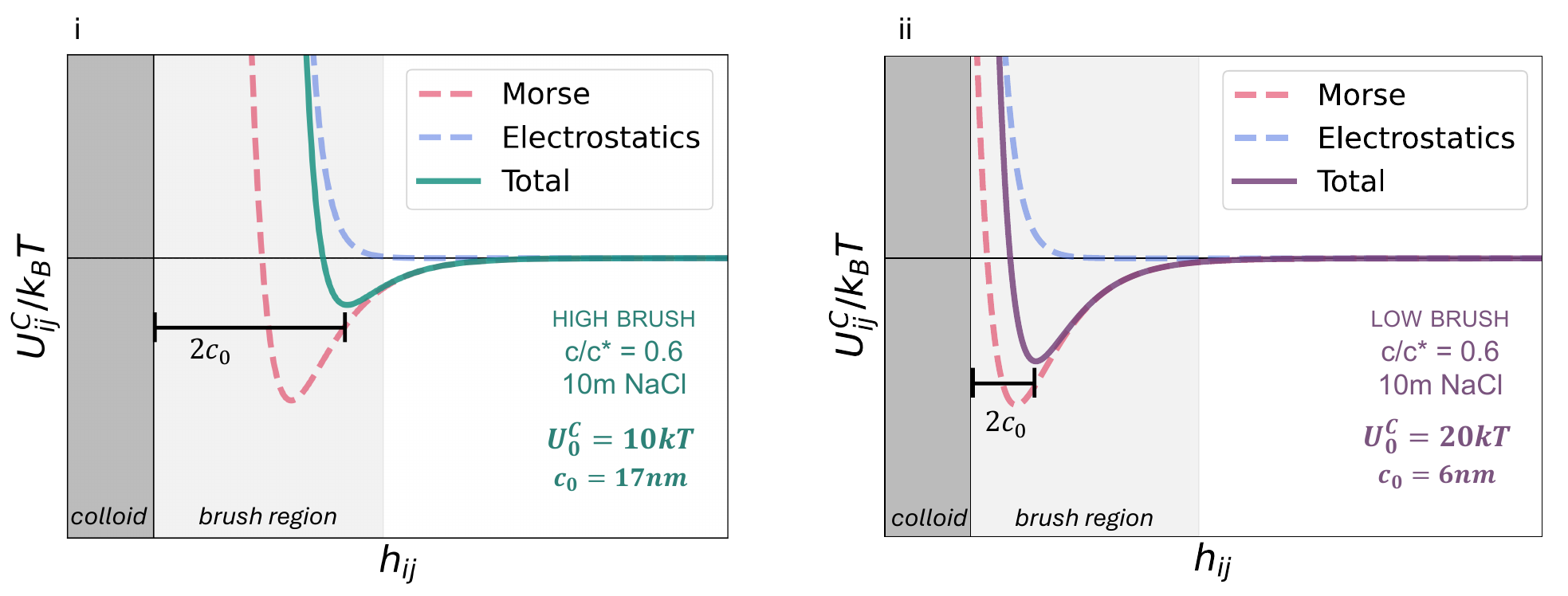}
\caption{\textbf{Summation of the total central-force interaction.} Shifted Morse depletion attraction (red) and electrostatic repulsion (blue) together produce a unique total potential for each combination of experimental parameters, with a total electro-steric shift $c_0$. The high brush density (i, green) and low brush density (ii, purple) systems experience very different interaction landscapes and different amounts of brush-brush interaction.}
\label{fig:SI_totalpotential}
\end{figure*}

In Eq. \ref{eq:doublelayer}, we introduce a brush-dependent offset distance $e_0$, representing changes in the location of the ``surface charge" due to reduced brush density. The Double Layer potential assumes a clear delineation between an ``effective electrostatic surface" and a surrounding ``diffuse ion layer". For a changing brush density, this picture is complicated by the distribution of surface charge across the brush region. Based on the chemistry of our brush, we expect particle charge to be distributed along most of the brush region rather than concentrated at either the bare colloid surface or the brush tip. Additionally, the small size of salt ions compared to our brush ($d_{ion}/h_b \approx 0.06$) and the curvature of our colloidal surface (which increases brush spacing near the brush tip) both suggest that the transition between an “effective electrostatic surface” and a surrounding “diffuse ion layer” should vary with changes in brush density and occur somewhere within the brush region (when $h_{ij}<2h_b$). The effect of electrostatic repulsion therefore depends on both the length of the effective charge distribution and the total charge of the brush.

\begin{figure*}[h!]
\centering
\includegraphics[width=1\linewidth]{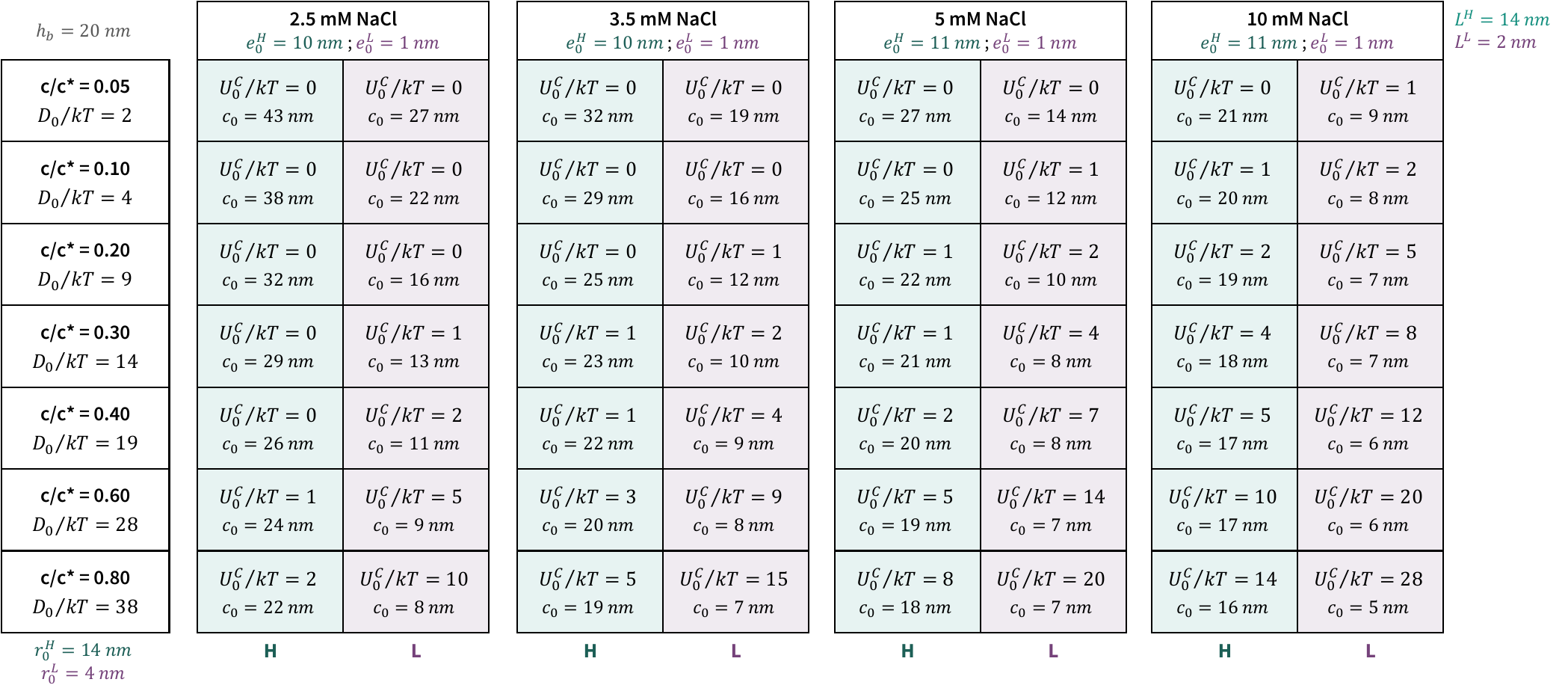}
\caption{\textbf{Simulation parameters corresponding to the full experimental phase space.} Depletion interactions are calculated from the attraction strength $D_0/kT$ and the brush density dependent steric minimum $r_0^H$ or $r_0^L$ using Eq. \ref{eq:morse}.  Changes in charge screening are calculated using the inverse Debye length (Eq. \ref{eq:kappa}) and a salt-dependent electrostatic surface $e_0^H$ or $e_0^L$, which is determined by fitting Equation \ref{eq:doublelayer} to a brush density dependent charge distribution. The depletion and electrostatic potential are summed together to produce a unique total pairwise potential $\mathbf{U}_{ij}^C$. Aggregation is controlled by the total attraction strength $U_0^C$. The degree of non-central constraints is controlled by $c_0$ and the strength of bending rigidity $K=U_0^C/kT$.}
\label{fig:table}
\end{figure*}

To approximate these effects with a single parameter, we select a per-particle charge layer thickness $L<h_b$ for each brush density: $L^H=14nm$ and $L^L=2nm$. We then distribute the surface charge, as calculated from experimental measurements of the Zeta potential, along this distance and recalculate a charge-distributed Double-Layer potential, shown in Fig.\ref{fig:SI_electrostatics}.i:
\begin{equation}
U_{ij}^{E,CD} (h_{ij}) = 
    \begin{cases}
        2 R_C \left( \frac{\pi \sigma^2}{\kappa^2 \varepsilon \varepsilon_0 (2L)} \right) e^{\left( \frac{\exp({\kappa h_{ij}})-1}{\kappa}\right)} & h_{ij}<2L \\
        2 R_C \left( \frac{\pi \sigma^2}{\kappa^2 \varepsilon \varepsilon_0 (2L)} \right) e^{\left( \frac{\exp({\kappa L})-1}{\kappa}\right)} & h_{ij}<2L \\
    \end{cases} 
    \label{eq:chargedistribution}
\end{equation}

After calculating the electrostatic repulsion for each system using Eq. \ref{eq:chargedistribution}, we opt to fit a traditional Double Layer potential to the charge-distributed model using Eq. \ref{eq:doublelayer}, as shown in Fig.\ref{fig:SI_electrostatics}.ii. This provides better numerical stability. It requires us to fit a different value of $e_0$ for each experimental salt concentration, but these unique values emerge naturally from the brush-dependent physics of our model. Together, the choice of $r_0$ and $e_0$ produce a unique total shift $c_0$ that determines the total interaction landscape, as shown in Fig.\ref{fig:SI_totalpotential}. The full list of parameters used to simulate each experimental composition is provided in Fig.\ref{fig:table}. Finally, non-central forces are included as an angle-based bending rigidity. When three or more particles form a cluster where their brush regions overlap, we record the initial three-body angle of this structure, $\theta_0$. At subsequent times, if the particle motion would change this angle, we apply a harmonic spring force, $\mathbf{U}^{NC}_{ijk}$, that maintains the current configuration:
\begin{equation}
    \mathbf{U}^{NC}_{ijk} (K,\theta) = 
        \begin{cases}
            \frac{1}{2} K (\delta \theta)^2 & \text{if } h_{ij} \leq 2h_b \\

            0 & \text{if } h_{ij} > 2h_b \\
        \end{cases} 
        \label{eq:rigidity}
\end{equation}
Here, $K$ is the bending stiffness in units of energy, scaled to be proportional to the magnitude of central force attraction at the potential minimum for each system: $K=U_{0}^C/k_BT$. The second parameter $\delta \theta = \theta_t - \theta_0$ is the amount of angular rearrangement that is being attempted at the current timestep, $t$, due to central forces. $\mathbf{U}^{NC}_{ijk}$ is applied as a tangential force. In this study $\mathbf{U}^{NC}_{ijk}$ is only applied when $h_{ij} \leq 2h_b$, directly relating the emergence of bending rigidity to interactions between neighboring brush regions. The effective range of rigidity also depends on the position of the total potential minimum $c_0$, such that $\Delta_{rigid}=2(h_b-c_0)$. In this way the model directly relates the emergence of bending rigidity to a reduction in brush density and changes in the effective electro-steric repulsion between particles. With both central and non-central forces calculated for all particles, the equations of motion then follows a typical Brownian dynamics formalism, as:

\begin{equation}
\mathbf{r}_i(t+\delta t) = \mathbf{r}_i(t) + \frac{\delta t}{6\pi \eta R_C}\left[ \left( \mathbf{F}_{i}^C + \mathbf{F}_{i}^{\text{bend}} \right) + \mathbf{F}_{i}^B \right]
\label{eq:particlemotion}
\end{equation}
where $\delta{t}$ is the integration time-step, $\eta$ is the dynamic fluid viscosity, $\mathbf{F}_{i}^C$ is the total central force per particle, $\mathbf{F}_{i}^{\text{bend}}$ is the total non-central force from bending rigidity per particle, and $\mathbf{F}_i^B$ is the Brownian force per particle. Additional hydrodynamic forces between particles are not considered.


\subsection{Justification of Particle Neighbor Search Distance}\label{search-distance}

To quantify gel microstructure from confocal image stacks, we identified particle neighbors based on a surface-to-surface search distance. For arrested (gelled) samples, a threshold of 0.4 {\textmu}m was applied. Although this distance is relatively large compared to the particle diameter (1.06 {\textmu}m), it accounts for experimental limitations such as particle motion during acquisition, limited spatial resolution especially in the \textit{z} direction, and microscope imaging noise, all of which can compromise reliable contact detection if a stricter criterion is used. This threshold provided robust coordination number statistics for arrested samples. However, when applied to diffusive suspensions, it yielded unphysically high values of ⟨Z⟩ ($\sim$1.4), which we attribute to amplified dynamic errors arising from particle motion over the relatively long imaging period ($\sim$2.5 min per stack). To mitigate this effect, a tighter cutoff of 0.2 {\textmu}m was adopted for diffusive samples, which more accurately reflects their instantaneous microstructure. These search distances (0.4 {\textmu}m for arrested samples and 0.2 {\textmu}m for diffusive samples) were used consistently throughout all experimental structural analyses presented in this work.

\begin{figure*}[h!]
\centering
\includegraphics[width=0.8\linewidth]{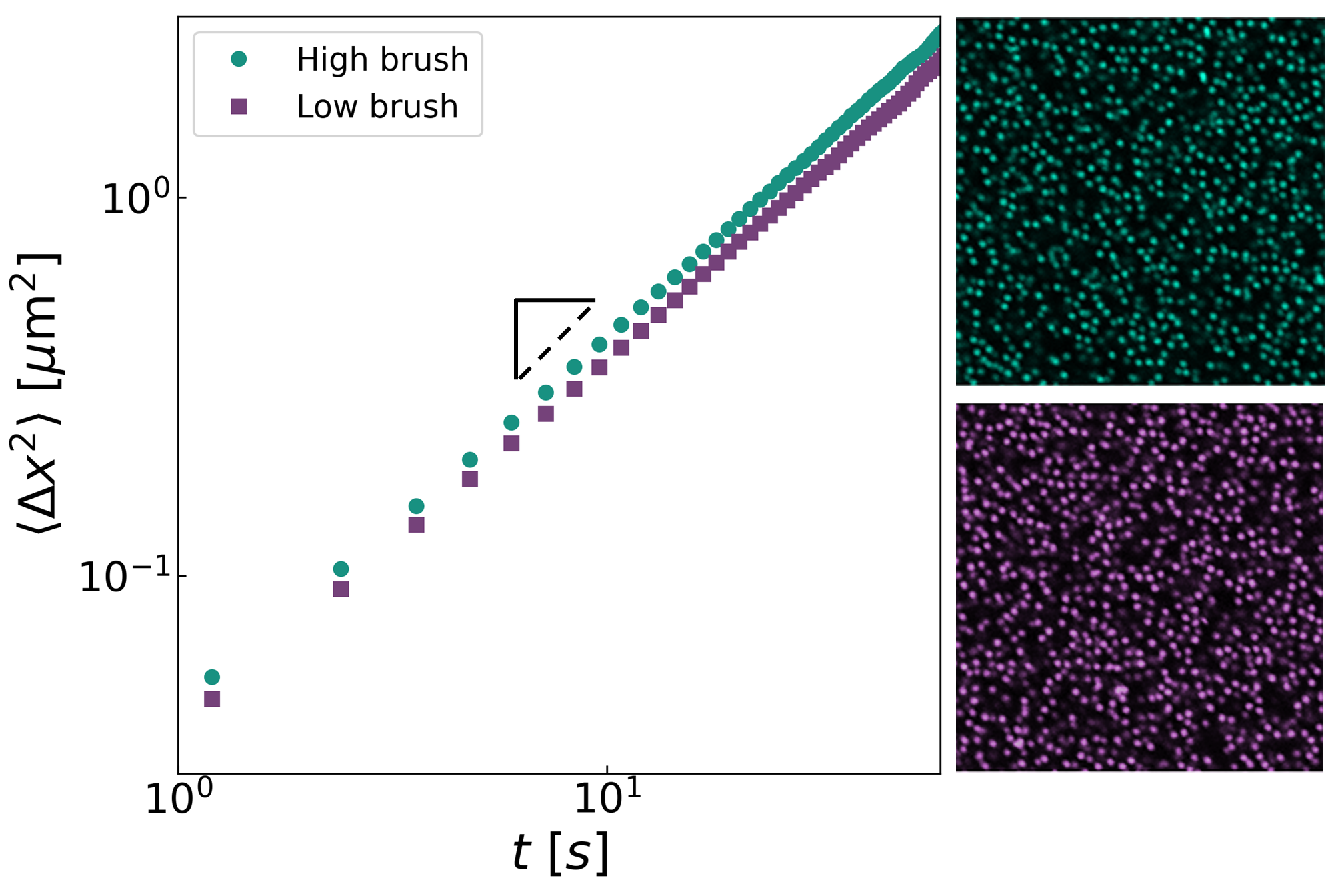}
\caption{\textbf{Charge screening baseline.} Mean-squared displacement (MSD) of high brush density (green) and low brush density (purple) colloidal suspensions in 10 mM NaCl without depletant. Both systems show diffusive dynamics with slope $\approx$ 1, consistent with freely dispersed particles. Confocal snapshots (right) confirm the absence of aggregation.}
\label{fig:noP}
\end{figure*}

\subsection{Control Experiments without Depletant}\label{baseline}

To confirm that the observed aggregation behavior across our mapping originates from depletion attraction rather than salt-driven effects, we performed control experiments using colloidal suspensions without any depletant polymer. High- and low brush density particles were each dispersed in solvent containing the highest NaCl concentration used in the main mapping (10 mM), corresponding to the strongest electrostatic screening condition. As shown in Fig.\ref{fig:noP}, both systems remained stable, as their mean-squared displacement (MSD) curves represent diffusive behaviors (slope $\approx$ 1). The corresponding confocal snapshots also reveal no signs of significant clustering for either brush density. These results confirm that even under the maximal ionic screening used in this work, there is no salt-induced aggregation between particles. Therefore, the attractive interactions responsible for network formation in this work arise solely from depletion forces induced by the non-adsorbing polymer.

\subsection{Simulation comparison without rigidity}\label{no-rigidity}

\begin{figure*}[h!]
\centering
\includegraphics[width=1\linewidth]{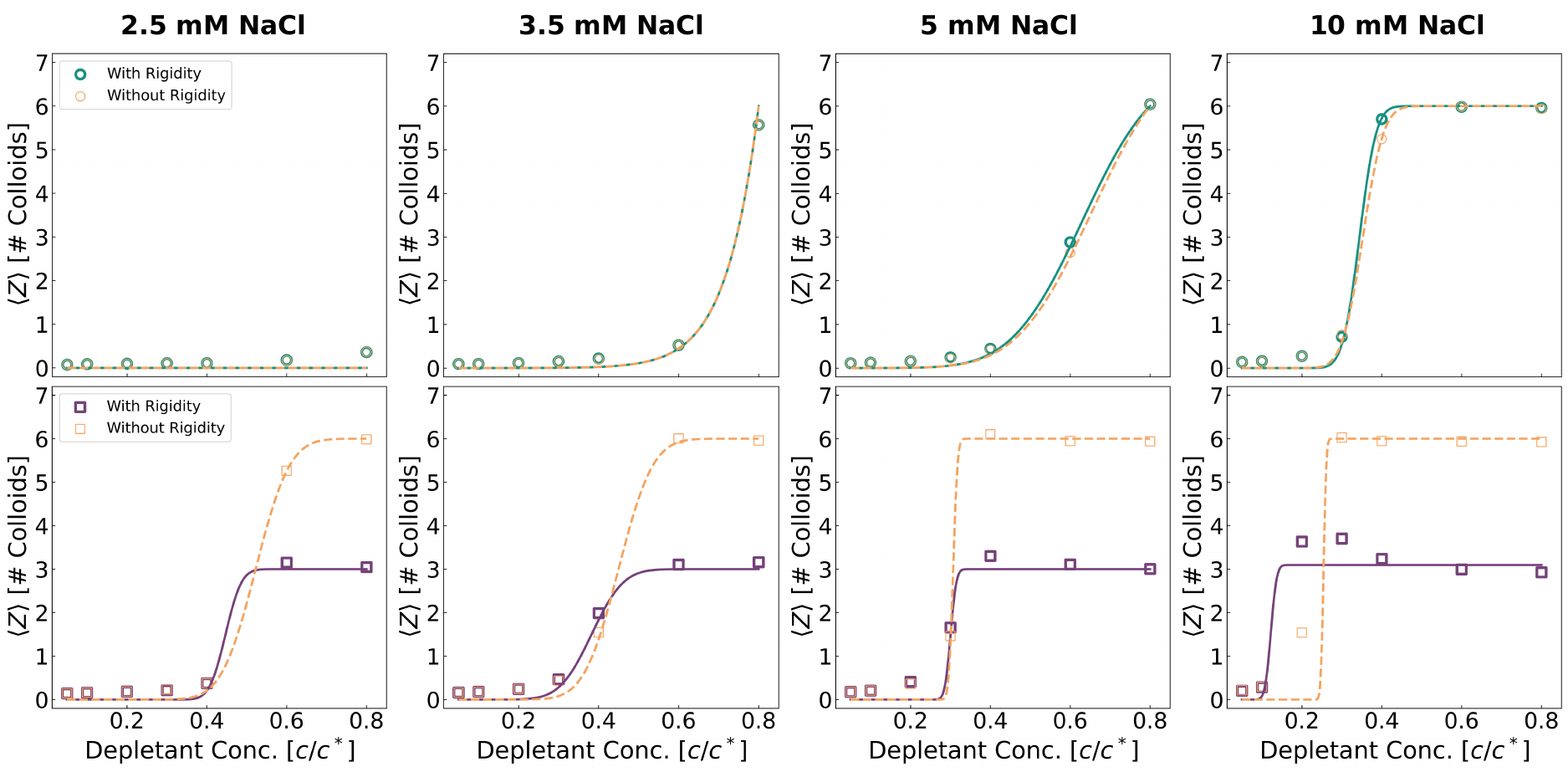}
\caption{\textbf{Comparison of simulation results with and without angular bending rigidity.} Average coordination number plotted across all simulated conditions with and without angular bending rigidity. Lines serve as a visual guide to the trends across depletant concentration. For high brush density systems (top), simulations without angular bending rigidity (orange, light symbols and dashed trends) produce nearly identical results to those with bending rigidity (green, bold symbols and solid trends). For low brush density systems (bottom), the gelation boundary is located at a similar depletant concentration in simulations with angular bending rigidity (purple, bold and solid) and without (orange, light and dashed); however, the microstructure is significantly different. These results demonstrate that the addition of non-central forces is essential for altering colloidal gel microstructure; however, changes in the location of the gelation boundary between brush densities depends more significantly on a change in electro-steric repulsion.}
\label{fig:SI_norigidity}
\end{figure*}

To isolate the effect of non-central forces we simulate the same total attraction potential with and without bending rigidity for all studied conditions. The average coordination number for these results is plotted in Fig.\ref{fig:SI_norigidity}.

For both high- and low brush density systems, the $\langle Z \rangle$ plateaus near the same experimental combination of depletant and salt with and without bending rigidity, reflecting a consistent gelation threshold. At very high screening in the low brush density system (10mM NaCl, bottom), angular bending rigidity does contribute a slight enhancement to gelation near the gelation boundary. Nevertheless, these results suggest that the majority of the change in position of the gelation boundary between the two brush densities is caused by a change in the central-force interaction landscape not the addition of non-central forces.

Conversely, angular bending rigidity has a significant effect on microstructure. Although the microstructure of the high brush density system is nearly identical with and without bending rigidity (reaching an $\langle Z \rangle \approx 6$), without bending rigidity the low brush density system also densifies to an $\langle Z \rangle \approx 6$. In contrast, the addition of non-central forces restricts the microstructure to a much lower $\langle Z \rangle \approx 3$, similar to the results seen in experiments (Fig.\ref{fig:full_data}).

\newpage

\subsection{Simulation gelation videos}\label{videos}

Time-series visualization of simulated gelation under high brush density (Video\_S1) and low brush density (Video\_S2) conditions mimicking experimental parameters of $c/c^*=0.6$ and 10mM NaCl. The high brush density system is dominated by central forces from the depletion interaction, forming traditional colloidal gel structures with thick strands of densely packed particles. Conversely, in the low brush density system the addition of non-central forces reduces the average coordination number, causing the system to quickly arrest into a ramified network of thin strands. Both networks remain stable over extended simulation times. By comparing the first few frames of video, you can also see that the high brush density system gels slightly slower than the low brush density case, as initial, more strand-like contacts coarsen into denser aggregates. This process does not appear in the low brush density simulation.

\subsection{Full Mapping of Microstructure Across All Conditions}\label{full-mapping}

\begin{figure*}[ht!]
\centering
\includegraphics[width=1\linewidth]{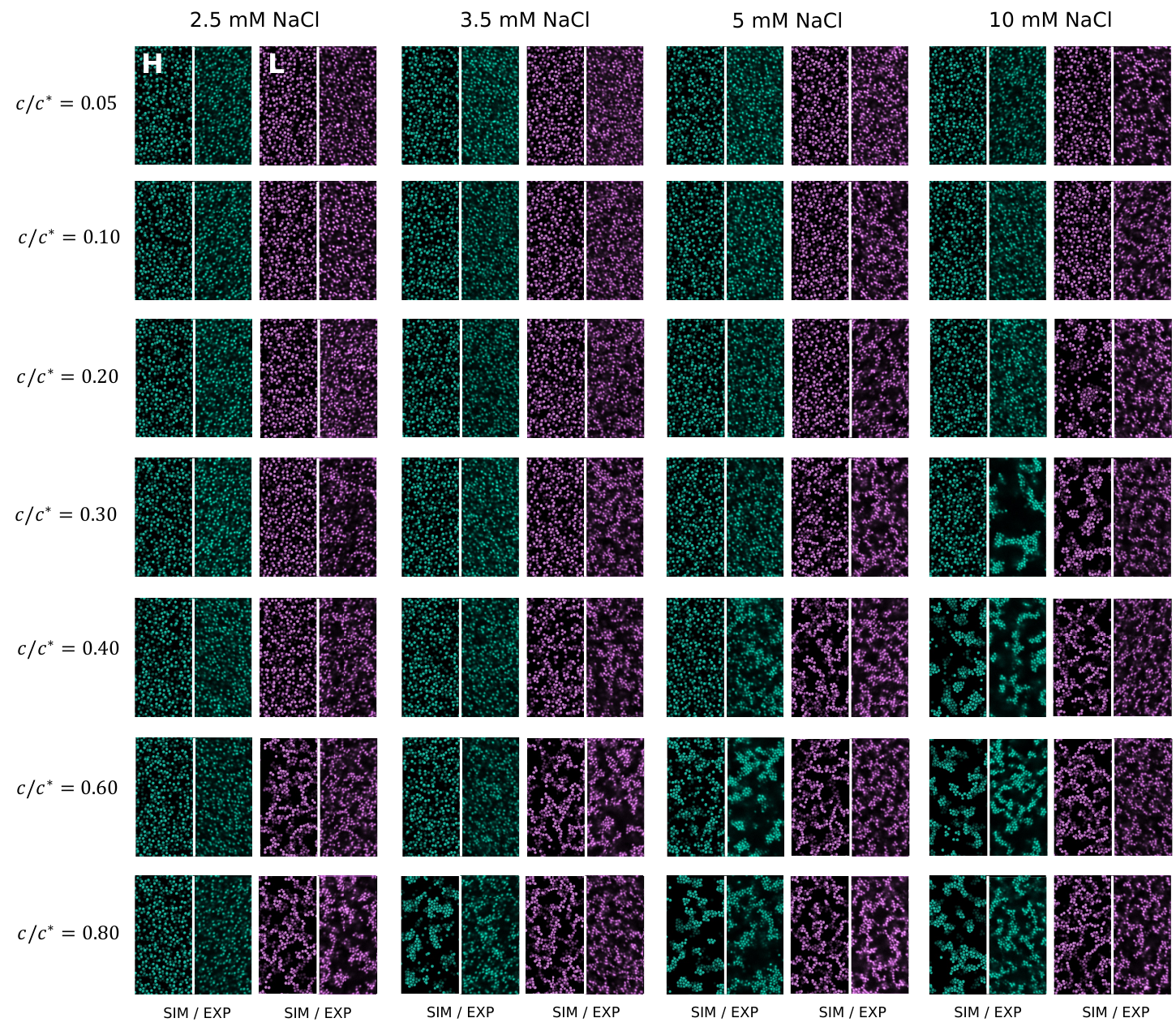}
\caption{\textbf{Full visual mapping.} Complete set of simulation (left) and confocal microscopy (right) images across all PAM and NaCl conditions for high (green) and low (purple)  brush density particles. Columns indicate increasing PAM concentration ($c/c^*=$ 0.05–0.8), and rows indicate increasing NaCl concentration (2.5–10 mM). The map illustrates the transition from dispersed to aggregated states and the influence of surface brush density on microstructure.}
\label{fig:SI_total_mapping}
\end{figure*}

Due to space limitations in the main text, only a selection of conditions was shown. Here in Fig.\ref{fig:SI_total_mapping}, we provide the complete dataset of colloidal structures across all tested polymer concentrations ($c/c^*=$ 0.05–0.8) and salt concentrations (2.5–10 mM NaCl), for high and low brush density particles. This includes paired simulation and confocal microscopy images for each condition, offering a comprehensive view of how depletion attraction and electrostatic screening influence network formation and microstructure under different brush densities.

\subsection{Quantitative Structural and Dynamical Analysis Across All Conditions}\label{full-data}

\begin{figure*}[ht!]
\centering
\includegraphics[width=1\linewidth]{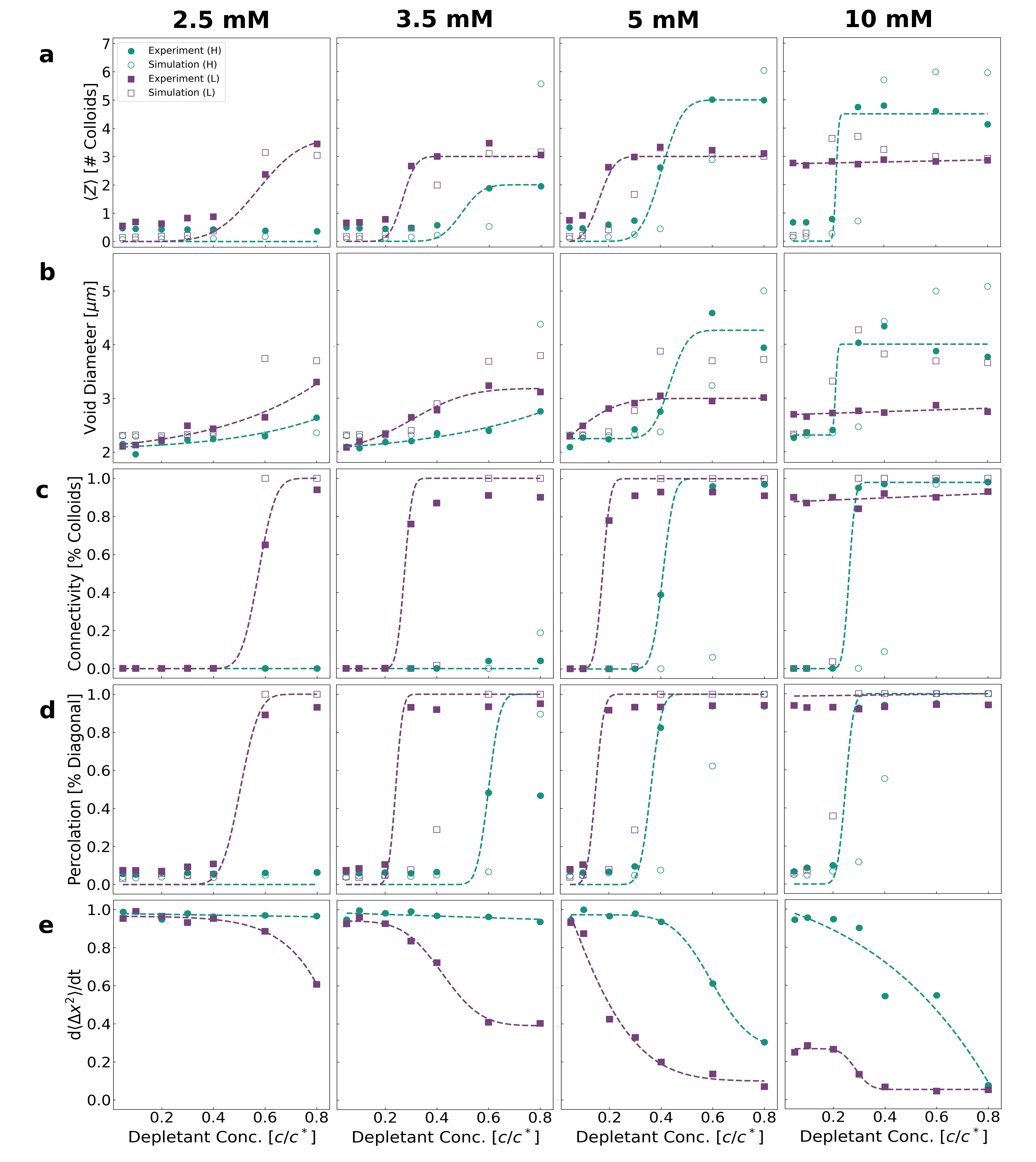}
\caption{\textbf{Full quantitative mapping. }Quantitative analysis of colloidal structure and dynamics with high (green) and low (purple) brush density particles across all salt concentrations. Columns represent increasing NaCl concentration (2.5–10 mM). a. Average contact number $\langle Z \rangle$; b. average void diameter; c. fraction of particles belonging to the largest connected component; d. percolation fraction defined as the percentage of the sample volume’s diagonal spanned by the largest cluster; e. slope of mean-squared displacement (MSD). Dashed curves are included as visual guides to highlight overall trends. Trends highlight key differences in packing, connectivity, and dynamic arrest between the two brush densities across a wide range of depletion strengths and charge screening conditions. }
\label{fig:full_data}
\end{figure*}

In the main text, we focused on the 5 mM NaCl case to illustrate structural differences between systems with high and low surface brush densities. Here, we present the complete dataset (Fig.\ref{fig:full_data}) across all salt concentrations (2.5, 3.5, 5, and 10 mM NaCl), showing key metrics extracted from both experiments and simulations. These include average contact number, average void diameter, largest connected component (LCC), percolation fraction defined by system-spanning network diameter, and mean-squared displacement slopes. Together, these results provide a comprehensive picture of how salt concentration and surface brush density jointly affect network formation and particle arrest.

\begin{figure*}[ht!]
\centering
\includegraphics[width=1\linewidth]{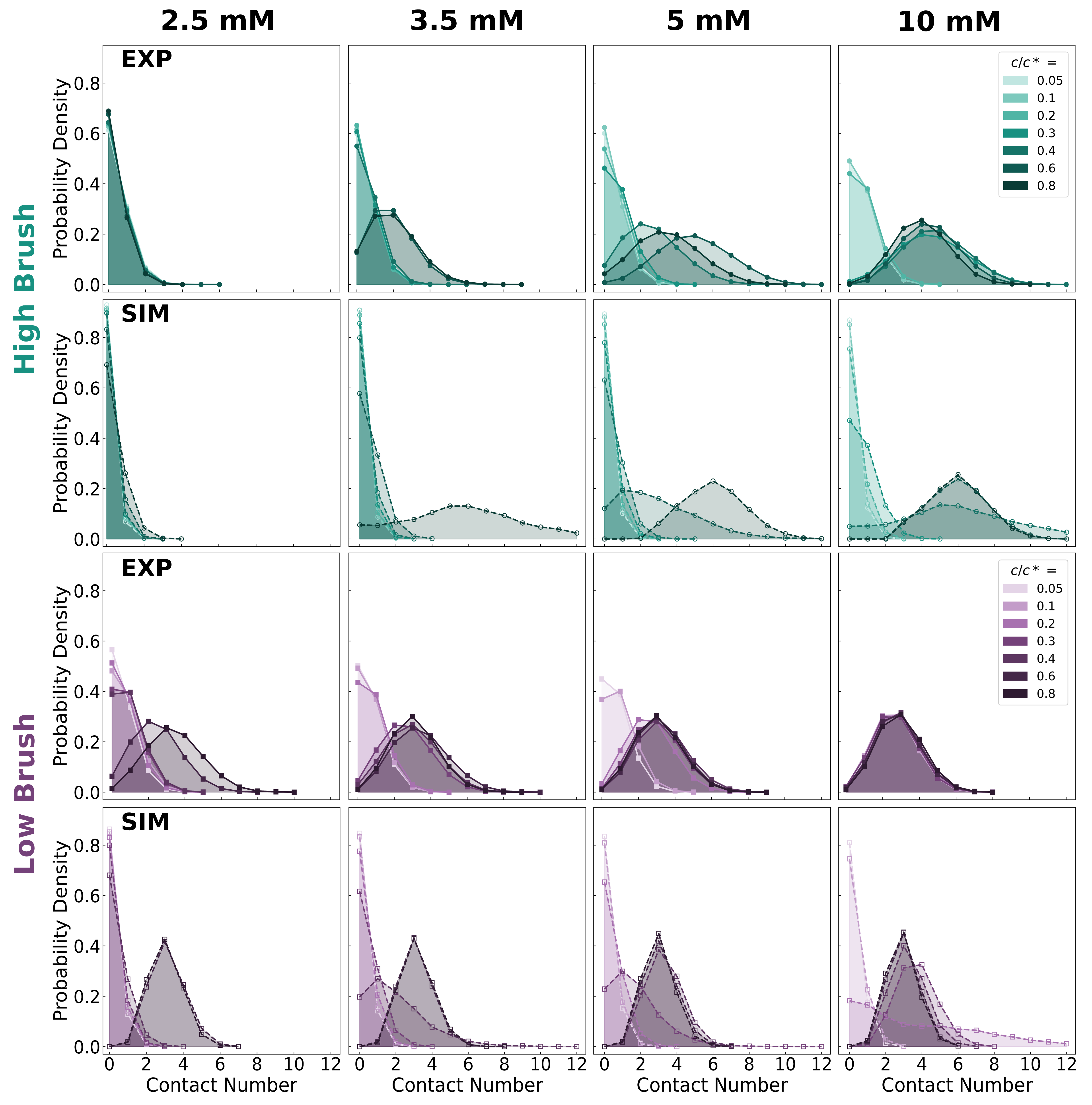}
\caption{\textbf{Contact number distributions from experiments and simulations across all PAM and salt concentrations.}}
\label{fig:contact_all}
\end{figure*}

\begin{figure*}[ht!]
\centering
\includegraphics[width=1\linewidth]{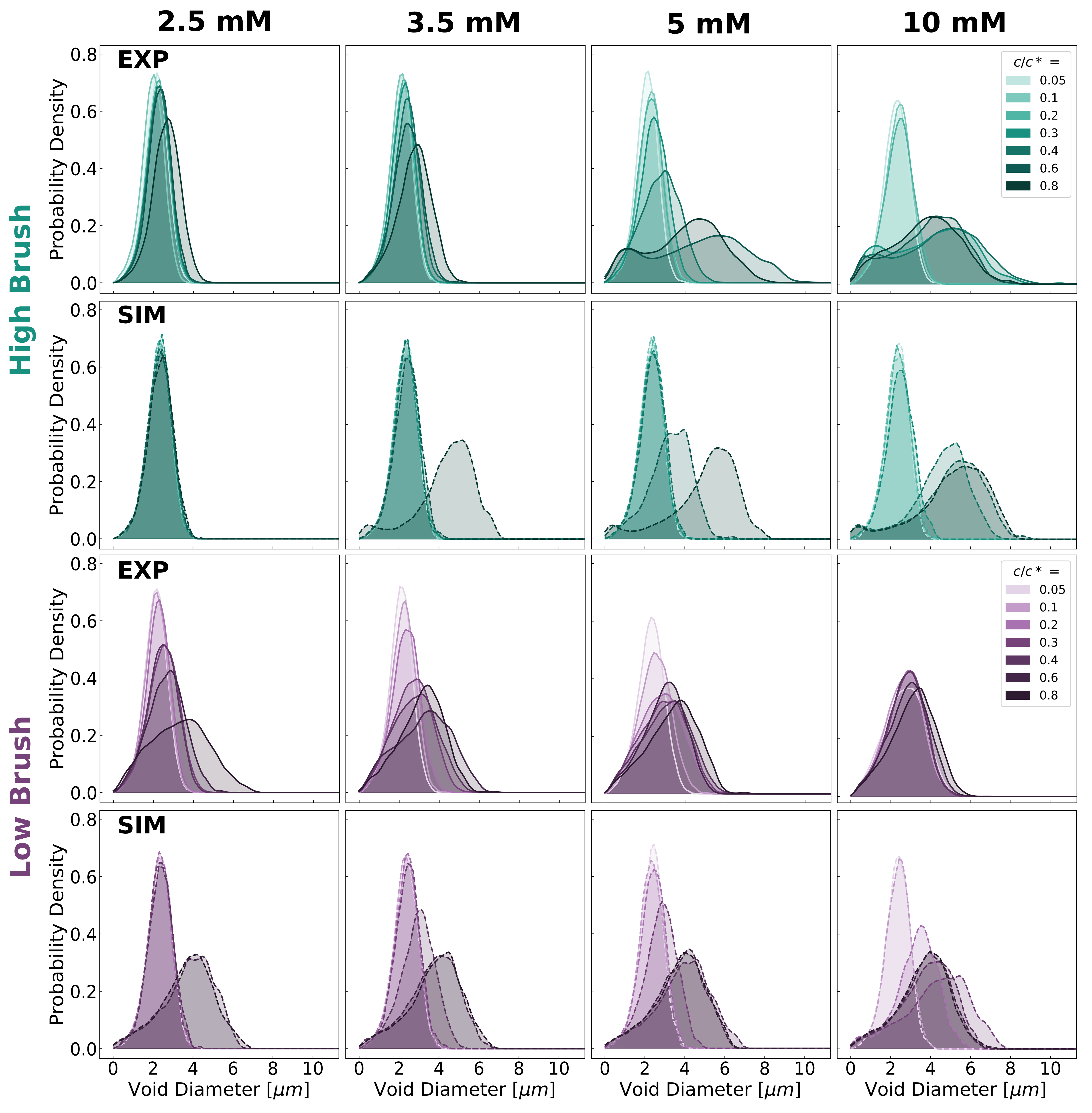}
\caption{\textbf{Void size distributions from experiments and simulations across all PAM and salt conditions.}}
\label{fig:voidsize_exp}
\end{figure*}

\begin{figure*}[ht!]
\centering
\includegraphics[width=1\linewidth]{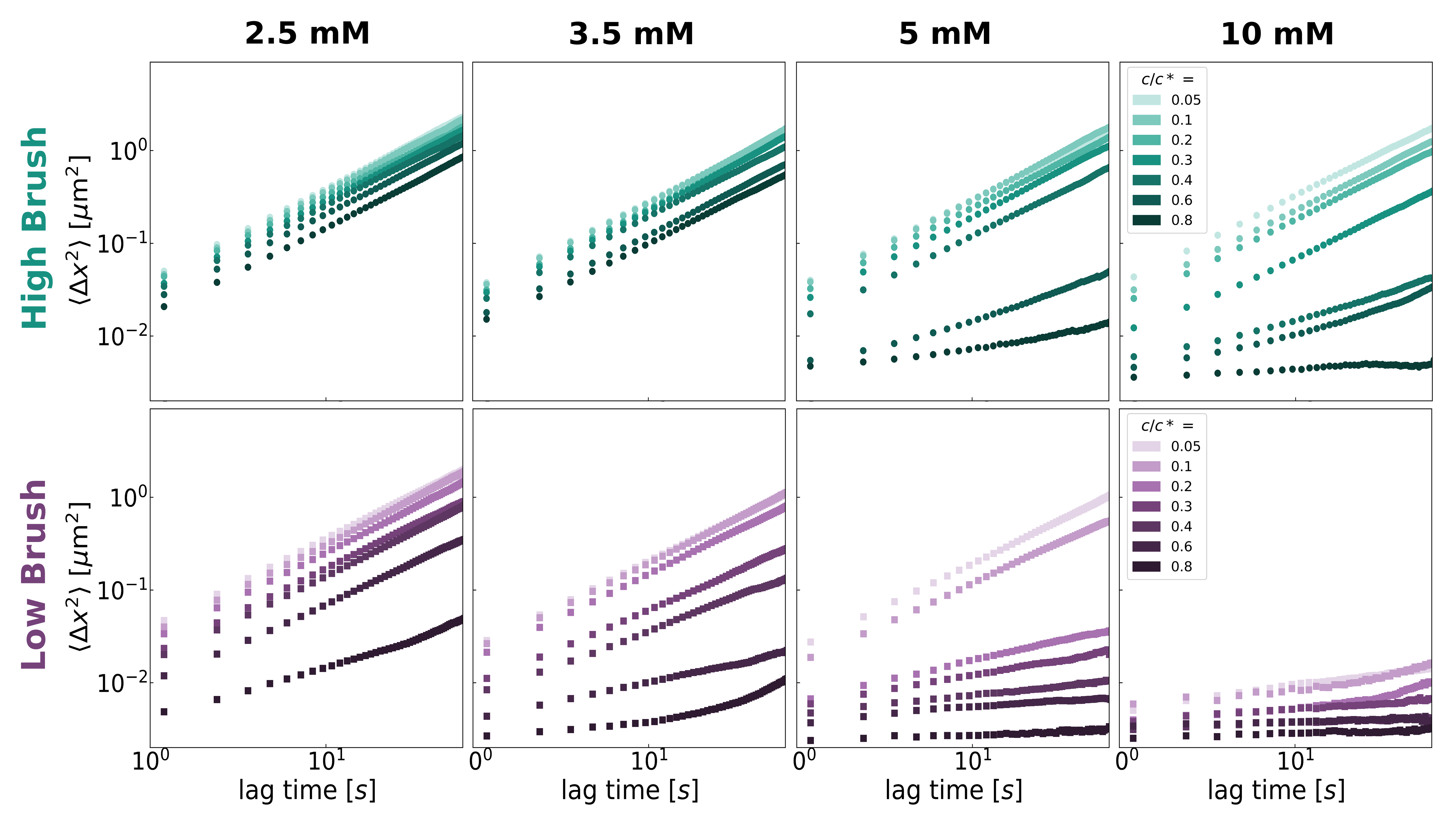}
\caption{\textbf{Mean-squared displacement (MSD) curves (experiments only) across all PAM and salt concentrations.}}
\label{fig:msd_SI}
\end{figure*}

The quantitative metrics reported in Fig.\ref{fig:full_data} were derived from underlying datasets shown here. Specifically, the average coordination numbers were calculated from the contact number distributions (Fig.\ref{fig:contact_all}), the average void sizes from the void width distributions (Fig.\ref{fig:voidsize_exp}), and the MSD slopes from the corresponding MSD curves (Fig.\ref{fig:msd_SI}). 

\begin{figure*}[ht!]
\centering
\includegraphics[width=0.8\linewidth]{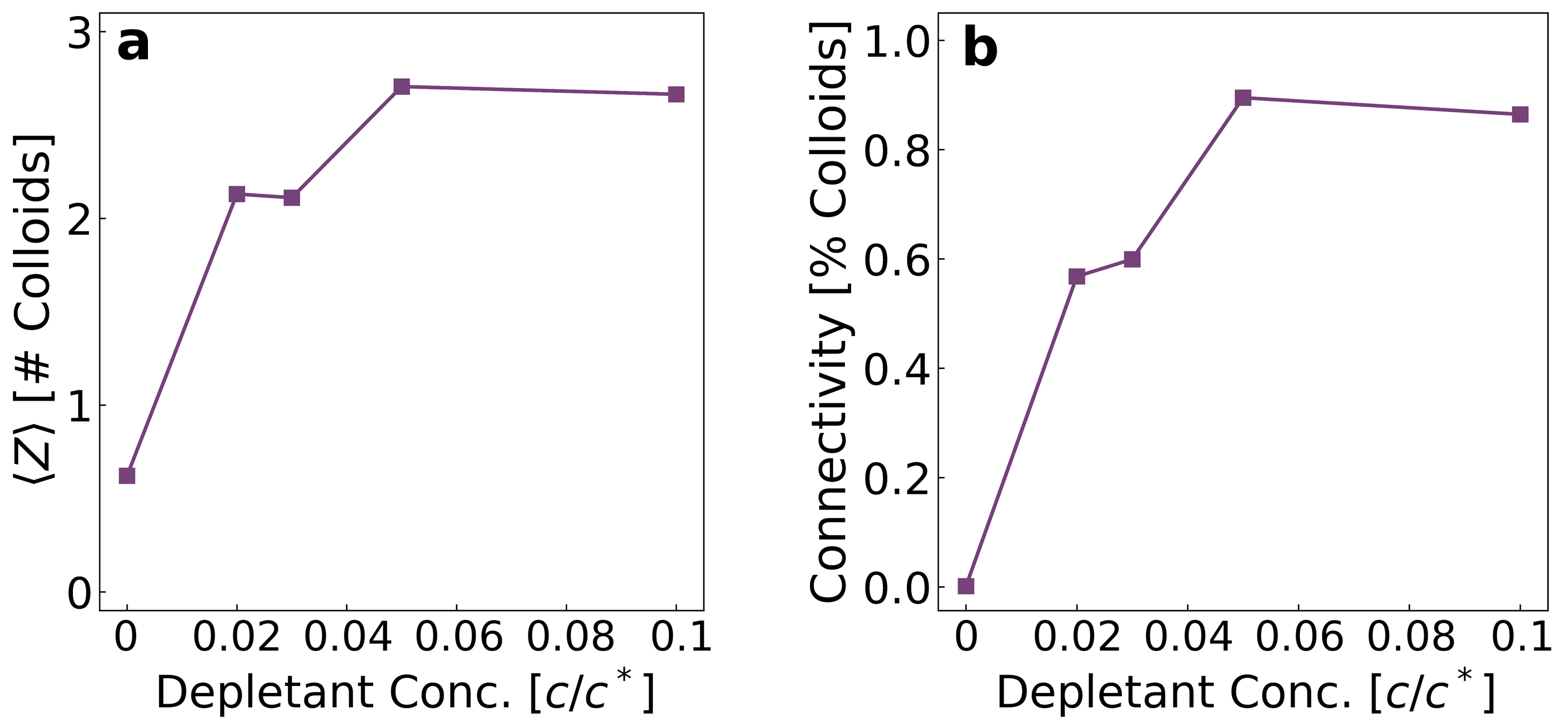}
\caption{\textbf{Gelation transition at 10 mM NaCl for low brush density systems.} (a) Average coordination number $\langle Z \rangle$ and (b) largest connected component (LCC) plotted as a function of PAM concentration, extending below the $c/c^*=0.05$ lower limit of the main mapping.}
\label{fig:transition}
\end{figure*}

For low brush density systems, the main mapping (Fig.\ref{fig:full_data}) shows clear transitions from dispersed to gelled states at intermediate salt concentrations (2.5–5 mM). However, at 10 mM NaCl, both the average coordination number $\langle Z \rangle$ and largest connected component (LCC) appear to plateau, with a gel state observed even at the lowest probed polymer concentration ($c/c^*=0.05$). To verify that this was not an artifact, we conducted additional measurements at PAM concentrations below $c/c^*=0.05$. As shown in Fig.\ref{fig:transition}, Both $\langle Z \rangle$ and LCC increase sharply between $c/c^*=0$ and 0.05 before reaching a plateau at higher concentrations, confirming that the gelation transition is still present but shifted to earlier conditions due to enhanced charge screening by salt. These results indicate that at high ionic strength, network formation occurs at significantly weaker depletion strengths compared to lower-salt conditions.

\clearpage

\makeatletter
\renewcommand{\@biblabel}[1]{[S#1]}
\makeatother
\putbib

\end{bibunit}

\end{document}